\documentclass[%
 reprint,
 amsmath,amssymb,
 aps,
]{revtex4-2}
\usepackage{enumitem}
\usepackage{color}
\usepackage{float}
\usepackage{graphicx}
\usepackage{subcaption}
\usepackage{tensor}
\usepackage{comment}
\usepackage{dcolumn}
\usepackage{bm}
\usepackage{hyperref}
\hypersetup{
	colorlinks=true,       
	linkcolor=blue,          
	citecolor=blue,        
	urlcolor=blue           
}
\usepackage{lipsum}
\usepackage{leftindex}

\newcommand*\diff{\mathop{}\!\mathrm{d}}

\begin{document}

\preprint{APS/123-QED}

\title{Axial quasi-normal modes of slowly rotating black holes in dynamical Chern-Simons gravity to second-order in spin and coupling}

\author{Tharaka Alapati}
\email{tharaka@iitb.ac.in}
\author{S. Shankaranarayanan}%
 \email{shanki@iitb.ac.in}
\affiliation{
Department of Physics, Indian Institute of Technology Bombay, Mumbai 400076, India 
}%
\date{\today}

\begin{abstract}
We compute the quasi-normal mode (QNM) frequencies of slowly rotating black holes in dynamical Chern-Simons (dCS) gravity, including corrections up to second order in the black hole's dimensionless spin parameter $\chi$ {(which is defined as  $J/M^2$, where $J, M$ are the angular momentum and mass of the black hole)} and second order in the dCS coupling parameter ($\alpha$). Due to the complexities of constructing a Newman-Penrose tetrad at this order, we employ a metric perturbation approach. We derive a system of coupled ordinary differential equations for the primary axial $l$-mode and the polar $l\pm 1$ modes, which is then solved numerically using the \emph{Runge–Kutta–Fehlberg method} with appropriate ingoing and outgoing wave boundary conditions. Our numerical framework is validated in the General Relativistic limit against known Schwarzschild QNMs and highly accurate Kerr QNM results for $\chi \leq 0.15$. For the fundamental $n=0, l=m=2$ axial mode, we present detailed numerical results illustrating the dependence of QNM frequencies on both $\chi$ and $\alpha$. We observe that while rotation generally increases the damping time, increasing the dCS coupling parameter significantly reduces the damping time of the axial mode. This finding contrasts with previous analytical work on polar modes, which suggested an increase in damping time due to dCS effects, highlighting a crucial parity-dependent difference in how dCS gravity impacts black hole ringdowns. Furthermore, we provide an analytical fitting formula for this mode: $M\omega = \omega_{00} + (\alpha/M^2)^2\omega_{02} + \chi(\omega_{10} + (\alpha/M^2)^2\omega_{12}) + \chi^2(\omega_{20} + (\alpha/M^2)^2\omega_{22})$. The coefficients $\omega_{00}, \omega_{10}$, and $\omega_{20}$ accurately reproduce Kerr QNM frequencies (in the limit $\alpha = 0$) for $\chi \leq 0.4$. These results, incorporating coupled spin and dCS effects at second order, provide more accurate theoretical predictions for testing dCS gravity with gravitational wave observations of black hole ringdowns. The refined QNM calculations are particularly relevant for lower-mass black hole merger events, such as GW230529, where dCS corrections may be more prominent and their distinct damping signatures could be observable by next-generation gravitational wave detectors.
\end{abstract}

\maketitle

\section{\label{sec:level1}Introduction}

The detection of gravitational waves (GWs) by the Laser Interferometer Gravitational-Wave Observatory (LIGO), Virgo and KAGRA  collaborations has opened a new era of GW astronomy, enabling stringent tests of gravity in the strong-field regime \cite{GW170104,2018-GWTC1-PRX,2020-GWTC2-PRX,KAGRA:2021vkt}. A key prediction of classical General Relativity (GR) is the \emph{no-hair theorem}, which states that stationary black holes (BHs) are fully characterized by three externally observable classical parameters: mass, charge, and angular momentum \cite{nohair,1987-Hawking.Israel-Book,1996-Heusler-Book,1998-Bekenstein-Arx,2012-Chrusciel.etal-LRR}. Current GW observations of binary BH mergers remain consistent with GR and this theorem~\cite{ToG,2021-ToG-GWTC2,2021-ToG-GWTC3}.  Nevertheless, probing potential deviations from GR remains a central goal in GW physics~\cite{2015-Berti.etal-CQG,2016-Cardoso.Gualtieri-CQG}, especially with next-generation detectors like Cosmic Explorer \cite{Evans:2023-CE} and the Einstein Telescope \cite{Branchesi:2023}, which will offer enhanced sensitivity and broader frequency coverage. These advancements may permit tighter constraints on BH properties, potentially revealing departures from the no-hair theorem that could signal modifications to GR~\cite{Barack:2018yly,Shankaranarayanan:2019yjx}.  

Dynamical Chern-Simons (dCS) gravity is a leading candidate for such modified gravity theories~\cite{dCS,Alexander:2009tp}. As the simplest parity-violating extension of GR, dCS gravity plays a pivotal role in constraining parity-violating gravitational effects~\cite{Alexander:2004us,Alexander:2004wk,Alexander:2024-amp,Chakraborty:2025qcu}. These effects arise from a non-minimal coupling between a dynamical scalar field and the Chern-Simons term of spacetime curvature, leading to BH solutions that violate the no-hair theorem~\cite{2007-Grumiller.Yunes-PRD,Yunes:2009hc,2021-Alexander.Yunes.etal-PRD,Okounkova:2019dfo,Okounkova:2021}. Specifically, in dCS gravity, BHs acquire additional structure through a scalar field dependent on the dCS coupling parameter~\cite{firstorderinspin,order2inspin}. While the Schwarzschild metric remains an exact solution, deriving exact rotating BH solutions in dCS gravity is challenging due to the complexity introduced by the Chern-Simons term and the scalar field~\cite{Stein:2014xba,Delsate:2018ome}. Consequently, analytical progress has relied on perturbative expansions in the spin parameter, typically truncated at second order~\cite{order2inspin}. These approximate solutions yield a modified metric that explicitly depends on the scalar field and dCS coupling, differing from the Kerr metric of GR. 

The GW emission from merging BHs consists of three phases: inspiral, merger, and ringdown~\cite{Sathyaprakash:2009xs,Maggiore,Creighton}. During the inspiral phase, where the orbital separation is large compared to the BH sizes, the system can be modeled using post-Newtonian (PN) theory~\cite{Blanchet2024-sh}. The GW amplitude and frequency increase monotonically until the merger phase, where the dynamics become highly nonlinear and short-lived~\cite{Creighton,Maggiore}. Following merger, the remnant BH relaxes to equilibrium through GW emission in the ringdown phase~\cite{REGGEWHEELER,ZERILLI}. This phase is characterized by a superposition of damped sinusoids, known as quasinormal modes (QNMs), whose frequencies are intrinsic to the BH and the underlying theory of gravity~\cite{VISHVESHWARA,Chandrasekhar,Nollert}.  

In GR, QNM frequencies and damping times are uniquely determined by the BH’s mass and spin and exhibit isospectrality --- identical spectra for odd- and even-parity perturbations~\cite{blackholesandbranesqnms,Kokkotas:1999bd,Konoplya:2011qq}. However, in modified theories like dCS gravity, the altered spacetime can break isospectrality~\cite{isospectralitybreaking} and shift the QNM spectrum~\cite{Manu,wagle,li2025perturbationsspinningblackholes,chung2025quasinormalmodefrequenciesgravitational}. While QNMs for first-order rotating dCS BHs have been computed using methods adapted from spherically symmetric cases~\cite{Manu,wagle}, extending these calculations to second-order spin solutions presents a significant analytical challenge.  

The computational challenges at second order in $\chi {(= J/M^2)}$ stem from the increasingly complex metric structure and strongly coupled perturbation equations inherent to higher-order rotational solutions. {Here, $J$ and $M$ refer to the angular momentum and mass of the BH, respectively.} While first-order analyses could safely neglect certain mode couplings, these effects become essential at $\mathcal{O}(\chi^2)$. The $l\leftrightarrow l\pm1$ couplings previously ignored at $\mathcal{O}(\chi)$ in \cite{Manu,wagle} now contribute significantly, while new $l\leftrightarrow l\pm2$ couplings emerge from quadrupolar spacetime deformations ($\propto M\chi^2$), higher-order frame-dragging, and dCS-modified rotational potentials. Attempts to decouple these equations via basis transformations ---successful in GR \cite{franchini2023slow} --- prove ineffective for dCS gravity due to the scalar field coupling, non-canonical potential terms, and parity-dependent mixing between perturbation sectors. This difficulty arises from the intrinsic coupling between the scalar field and metric perturbations, the presence of non-canonical potential terms, and parity-dependent mixing between the perturbation sectors. In fact, even at linear order in $\chi$, completely decoupling the axial perturbations from the scalar field perturbation equations is not possible.  Hence, in Ref.~\cite{Manu}, only approximate axial QNM frequencies, which ignored this coupling, were reported.

In this work, we tackle these challenges by numerically integrating a \emph{reduced set of eight coupled ordinary differential equations} to obtain the axial QNM frequencies at second order in $\chi$. It's important to emphasize that all calculations leading up to this numerical integration step are analytical and exact. This approach contrasts with the recent work by Chung et al. \cite{chung2025quasinormalmodefrequenciesgravitational}, which employed purely numerical methods, validating their QNM frequencies for $\chi \leq 0.75$ at $\mathcal{O}(\alpha^2)$. Consequently, our work offers a complementary perspective to existing numerical methods in the literature, both at linear and higher orders in $\chi$ \cite{wagle,chung2025quasinormalmodefrequenciesgravitational}. Specifically, this work bridges these different computational regimes by semi-analytically computing the fundamental ($n=0$), $l=m=2$ axial QNM frequency to $\mathcal{O}(\chi^2,\alpha^2)$ precision, making it valid for $\chi \leq 0.15$ and $\alpha/M^2 \leq 0.05$. We also derive an analytic fitting formula for the axial QNM frequency. This derived formula achieves sub-percent accuracy for $\chi \leq 0.4$, thus providing a practical and efficient tool for gravitational waveform modeling in dCS gravity. Our results show good agreement with those presented in Ref.~\cite{chung2025quasinormalmodefrequenciesgravitational}.

In principle, our analytical framework can be straightforwardly extended to compute the polar QNM frequencies, which would necessitate solving a \emph{reduced set of ten coupled ordinary differential equations}. However, the dCS correction to the polar QNM frequency is considerably smaller than that for the axial mode, primarily because the polar mode lacks a leading-order in spin correction unlike the axial mode. Therefore, from a computational necessity standpoint, evaluating the polar QNM frequencies is not pursued in this work.

We use $(-,+,+,+)$ signature for the 4-D spacetime metric~\cite{Gravitationbook}. We use the geometric units $G=c=1$ and $\kappa=1/(16\pi)$. Without loss of generality, we set one of the dCS parameter $(\beta)$ to unity. Our convention for the Levi-Civita tensor is $\epsilon^{1234}=+1/\sqrt{-g}$. The various physical quantities with the \emph{over-line} refers to the values evaluated for the background space-time.

\section{Slowly Rotating BH Solution in dCS gravity}
\subsection{Dynamical Chern-Simons (dCS) gravity}

As mentioned in the introduction, dCS gravity is a four-dimensional, parity-violating theory wherein a dynamical pseudoscalar field $\vartheta$ is non-minimally coupled to the spacetime curvature via the Pontryagin density \cite{dCS,Alexander:2009tp}. The vacuum dCS action is given by:
\begin{align}
S:=S_\text{EH}+S_\text{CS}+S_{\vartheta} 
\end{align}
where, $S_\text{EH}$ is the standard Einstein-Hilbert action, 
$S_\vartheta$ describes the dynamics of the pseudoscalar field and 
$S_\text{CS}$ introduces the parity-violating coupling:
\begin{align}
      S_\text{EH}&=\kappa\int_{\mathcal{M}} d^4x\,\sqrt{-g}R\\
    S_\vartheta&=-\frac{\beta}{2}\int_{\mathcal{M}} d^4x\,\sqrt{-g}\big[g^{ab}\nabla_{a}(\vartheta)\nabla_{b}(\vartheta)+2V(\vartheta)\big]\\
    S_\text{CS}&=\frac{\alpha}{4}\int_{\mathcal{M}} d^4x\,\sqrt{-g}\vartheta\leftindex{}^{*}RR
\end{align}
The parity-violating term, $\leftindex{}^{*}RR$, is proportional to the wedge product $R \wedge R$ and is known as the Pontryagin density. It is explicitly defined as:
\begin{equation}
\leftindex{}^{*}RR=\leftindex{}^{*}\tensor{R}{^a_b^{cd}}\tensor{R}{^b_{acd}}\end{equation}
where, the dual Riemann tensor is defined as,
\begin{equation}
\leftindex{}^{*}\tensor{R}{^a_b^{cd}}=\frac{1}{2}\epsilon^{cdef}\tensor{R}{^{a}_{bef}}
\end{equation}
with $\epsilon^{cdef}$ being the $4-D$ Levi-Civita tensor.  Due to the dynamical nature of the pseudoscalar field $\vartheta$, this is referred to as dynamical Chern Simons. In $\vartheta=0$ limit, dCS gravity precisely reduces to General Relativity. We set $\beta=1$ without loss of generality.

Varying the total action with respect to the metric $g_{ab}$ and the scalar field $\vartheta$ yields the dCS field equations:
\begin{equation}
\begin{aligned}
G_{ab} + \frac{\alpha}{\kappa}C_{ab} &= \frac{1}{2\kappa}T_{ab}^{\vartheta} \\
\Box\vartheta &= \frac{dV}{d\vartheta} - \frac{\alpha}{4}\leftindex{}^{*}RR
\end{aligned}
\label{eq:dCS-fieldeq}
\end{equation}
where, the Cotton tensor ($C_{ab}$) is given by:
\begin{equation}\label{C-tensor}
    C^{ab}=\vartheta_{c}\epsilon^{cde(a}\nabla_{e}R^{b)}\,_{d}+\vartheta_{cd}\leftindex^{*}R^{d(ab)c}
\end{equation}
and the parentheses denote symmetrization of indices, and $\vartheta_{c}=\nabla_{c}\vartheta$ and $\vartheta_{cd}=\nabla_{c}\nabla_{d}\vartheta$.

$T^{\vartheta}_{ab}$ represents the stress-tensor of the pseudoscalar field:
\begin{equation}
    T_{ab}^\vartheta= \Big[(\nabla_a\vartheta)(\nabla_b\vartheta)-\frac{1}{2}g_{ab}(\nabla_c\vartheta)(\nabla^c\vartheta)-g_{ab}V(\vartheta)\Big]
\end{equation}
  Since we are interested in the study of QNMs of BHs in dCS, which are characteristic modes of a BH in vacuum, we assume the absence of external matter fields and also set the pseudoscalar potential $V(\vartheta)$ to zero.

It is important to note that the Pontryagin density, $\leftindex{}^{*}RR$, vanishes for any spherically symmetric metric. Consequently, dCS gravity reduces to standard GR for such configurations, implying that the Schwarzschild metric is also a vacuum solution of dCS gravity~\cite{dCS}. However, for axisymmetric spacetimes, the Pontryagin density does not vanish. Thus, the Kerr metric, a stationary axisymmetric solution of GR, is not a solution in dCS theory~\cite{Yunes:2009hc}. Obtaining an exact rotating BH solution in dCS gravity is challenging due to the coupled, non-linear nature of the field equations. Therefore, slowly rotating solutions have been derived perturbatively in the spin parameter $\chi$ and the Chern-Simons coupling parameter $\alpha$, with the most accurate analytical solutions currently available up to $\mathcal{O}(\chi^2,\alpha^2)$~\cite{order2inspin}. 

\subsection{Slowly rotating BH solutions in dCS}

{The background metric for a slowly rotating BH in dCS gravity, expanded to $\mathcal{O}(\chi^2)$ and $\mathcal{O}(\alpha^2)$, has been a subject of detailed study \cite{Yunes:2009hc, Konno10.1143/PTP.122.561, Pani:2011xj}. 
In the limit where the dCS coupling parameter $\alpha$ vanishes, the dCS BH solution must smoothly recover the Kerr metric of GR.} 

Utilizing BH perturbation theory, the dCS corrections to the slowly rotating BH metric have been derived up to $\mathcal{O}(\chi^2,\alpha^2)$, where $\chi$ is the dimensionless spin parameter \cite{order2inspin}. Notably, the metric component $g_{t\phi}$ and the pseudoscalar field $\vartheta$ exhibit corrections only at linear order in the spin parameter, specifically $\mathcal{O}(\chi,\alpha^2)$. In contrast, other metric components, including $g_{tt}$, $g_{rr}$, $g_{\theta\theta}$, and $g_{\phi\phi}$, acquire corrections at second order in both $\chi$ and $\alpha$, i.e., $\mathcal{O}(\chi^2,\alpha^2)$. The total metric can be expressed as:
\begin{align}
    ds^2&=ds_K^2+\chi ds_{\text{dCS1}}^2+\chi^2ds_{\text{dCS2}}^2
\end{align}
where, $ds_K^2$ is the Kerr line element and $ds_{\text{dCS1}}^2$ is given by,
\begin{align}
    ds_{\text{dCS1}}^2&=\frac{5}{4}\frac{\alpha^2}{\kappa}\frac{M}{r^4}\Bigg(1+\frac{12}{7}\frac{M}{r}+\frac{27}{10}\frac{M^2}{r^2}\Bigg)dtd\phi
\end{align}
The full expression for $ds_{\text{dCS2}}^2$, encompassing the second-order corrections, is detailed in Appendix \ref{Appendix_A}. The background pseudoscalar field ($\bar{\vartheta}$) itself also acquires a perturbation due to rotation, appearing at $\mathcal{O}(\chi)$ relative to the non-rotating solution:
\begin{equation}
    \bar{\vartheta}=-\frac{5}{8}\chi\alpha\frac{\cos\theta}{r^2}\Bigg(1+\frac{2M}{r}+\frac{18M^2}{5r^2}\Bigg) \, .
\end{equation}
The event horizon of this slowly rotating dCS BH is determined by finding the largest real root of the equation $g_{t\phi}^2 - g_{tt}g_{\phi\phi} = 0$ \cite{Poisson:2009pwt}. At first order in $\chi$, the horizon radius remains at $r_H = 2M$, consistent with the non-rotating Schwarzschild solution. However, considering corrections up to $\mathcal{O}(\chi^2,\alpha^2)$, the horizon radius for this slowly rotating dCS BH is modified to \cite{order2inspin}:
\begin{equation}
\label{eq:horizondCS}
    r_H=2M-\frac{\chi^2M}{2}\Bigg(1+\frac{915\alpha^2}{14336\kappa M^4}\Bigg)
\end{equation}
This expression clearly shows how the dCS coupling parameter $\alpha$ influences the horizon radius, deviating from the GR prediction of $r_H = 2M(1-\chi^2/4)$ for a slowly rotating Kerr BH.

Constraints on the dCS coupling constant have been derived from various astrophysical observations. Particularly relevant are the constraints obtained from the X-ray observations of accreting BHs. For instance, the latest analysis by Silva et al. using NICER (Neutron Star Interior Composition Explorer) data from the X-ray binary system MAXI J1803-298 yielded a stringent constraint on the dCS coupling constant \cite{NICER1}:
$$\sqrt{\alpha} \leq 8.5 \, \text{km}$$
Converting this constraint into a dimensionless form relevant for BH studies, we obtain:
$$\frac{\alpha}{M^2} \leq 33.11 \left(\frac{M_\odot}{M}\right)^2$$
where $M_\odot$ is the solar mass. To appreciate the potential impact of dCS corrections, consider a BH with a mass of $10 \, M_{\odot}$. For such a BH, the ratio of the dCS correction term to the purely GR correction in the event horizon radius $r_H$ is given by: 
$$\frac{915\alpha^2}{14336 \kappa M^4} \leq 1.06$$
This result is crucial as it indicates that the dCS correction to the horizon radius is of a comparable magnitude to the purely GR spin-squared correction. This non-negligible contribution to a fundamental BH property strongly suggests that the QNM corrections arising from dCS gravity will be significant, particularly for BHs in the lower mass range, such as those around $\sim 10\,M_{\odot}$. This highlights the importance of incorporating second-order spin corrections in our analysis to accurately capture the phenomenology of dCS BHs and their potential observational signatures.

{Before proceeding with the perturbation analysis, it is crucial to address the definition of the spin parameter, $\chi$, for a BH in a modified theory of gravity. For a generic non-GR BH, identifying parameters that correspond directly to the mass and angular momentum of the Kerr solution can be a subtle issue.

In this work, we ground our definitions in the fundamental, conserved charges of the asymptotically flat spacetime. The parameters $M$ and $J$ correspond to the Arnowitt-Deser-Misner (ADM) mass and ADM angular momentum, respectively~\cite{Poisson:2009pwt}. These quantities are calculated from surface integrals at spatial infinity and represent the total mass-energy and angular momentum of the system as perceived by a distant observer. The dimensionless spin parameter is then defined unambiguously as ${\chi \equiv J/M^2}$.

This definition ensures a direct and clear perturbative connection to the Kerr solution~\cite{firstorderinspin}. The dynamical Chern-Simons theory possesses a smooth limit to GR as the coupling constant $\alpha \to 0$. In this limit, the dCS BH solution reduces to the Kerr solution, and its ADM mass $M$ and ADM angular momentum $J$ become identical to the corresponding Kerr parameters. Therefore, the parameter $\chi$ used throughout our analysis is not an arbitrary choice but represents the same physical quantity as the spin of a Kerr BH.

Our methodology consists of fixing these asymptotic observables ($M$ and $J$) and then calculating how the modifications to the near-horizon geometry induced by the non-zero dCS coupling $\alpha$ affect the quasi-normal mode spectrum. The resulting frequencies, $\omega(\chi, \alpha)$, are thus the characteristic frequencies of a dCS BH that has the same total mass and angular momentum as a Kerr BH of spin $\chi$ but whose structure is modified by the 'hair' associated with the dCS pseudoscalar field.}

\section{Linear Perturbations in DCS gravity}

As mentioned earlier, Srivastava et al~\cite{Manu} analytically obtained QNMs for slowly rotating BHs in dCS gravity up to first order in spin. However, extending this analysis to second order in rotation ($\chi^2$) and second order in the dCS coupling parameter ($\alpha^2$) necessitates a distinct and more involved methodology.

\subsection{Challenges at second order in spin}

The primary challenge arises from the structure of the spacetime metric itself. For a slowly rotating BH at the first order in spin ($\chi$), 
the metric corrections are relatively contained, primarily affecting the $g_{t\phi}$ component, which introduces frame-dragging. However, at second order in spin ($\mathcal{O}(\chi^2)$), the metric becomes considerably more complex. Corrections appear in multiple components, including the diagonal terms $g_{tt}, g_{rr}, g_{\theta\theta}, g_{\phi\phi}$, as well as  $g_{t\phi}$, depending on the dCS contributions.

It is common practice to derive QNMs using the Teukolsky equation within the Newman-Penrose (NP) formalism, especially for rotating BHs in GR~\cite{Teukolsky:1973ha}. However, constructing the appropriate null tetrad for the dCS metric at second order in spin and consistently incorporating dCS corrections into the NP curvature components (like $\Psi_0$ or $\Psi_4$) becomes highly non-trivial~\cite{Berti:2005eb}. The dCS corrections modify the NP equations themselves in a complex manner, making the separation of variables, if at all possible, a formidable task. Hence, in this work, we adopt the \emph{metric perturbation method}~\cite{Pani,Manu,Pierini:2021jxd,edgbqnm2}.

The metric perturbation method, while algebraically intensive, offers a more direct route when the symmetries exploited by the NP formalism are obscured by higher-order corrections or theory modifications. However, the procedure followed by Srivastava et al. \cite{Manu} is likely insufficient here primarily due to the intricate coupling of perturbation equations induced by the $\mathcal{O}(\chi^2)$ terms in the background metric and the dCS modifications. The dCS corrections often lead to parity-violating effects and mode couplings that are absent or simpler in GR. Our proposed procedure is as follows: \\[2pt]

\noindent 1.~{\bf Metric perturbations in the Regge-Wheeler gauge and obtain the perturbation equations:}
~We start by perturbing the dCS corrected slowly rotating BH metric  ($\bar{g}_{\mu\nu}$) as 
\begin{equation}\label{pert_metric}
    g_{\mu\nu}=\bar{g}_{\mu\nu}+\epsilon\, h_{\mu\nu}\quad\quad\text{and}\quad\quad\vartheta=\bar{\vartheta}+\epsilon\,\delta\vartheta
\end{equation}
where $\epsilon$ is a book keeping parameter, $h_{\mu\nu}$ and $\delta \vartheta$ are perturbations. 
The Regge-Wheeler gauge~\cite{REGGEWHEELER,Nollert} is a convenient choice for handling perturbations of spherically symmetric backgrounds, and its generalization is adapted here \eqref{hmunu}. For rotating backgrounds, one typically separates perturbations into axial (odd-parity) and polar (even-parity) sectors. In dCS gravity, however, the parity symmetry can be broken by the $\vartheta$ field, leading to coupling between these sectors and the scalar field~\cite{Pani:2011xj}. The linearized field equations are:
\begin{eqnarray}
\delta G_{\mu\nu} + \frac{\alpha}{\kappa} \delta C_{\mu\nu} &=& \frac{1}{2\kappa}\delta T_{\mu\nu}^{\vartheta} \\ 
\overline{\Box} \delta\vartheta + \delta(\Box) \vartheta^{(0)} &=& - \frac{\alpha}{4} \delta(\leftindex{}^{*}RR)
\end{eqnarray}
These equations will contain terms up to $\mathcal{O}(\chi^2)$ from the background and will be solved for $h_{\mu\nu}$ and $\delta\vartheta$, keeping terms consistent with $\mathcal{O}(\alpha^2)$ frequency corrections. We represent the modified Einstein equations by $\delta E_{\mu\nu}$,
\begin{equation}\delta E_{\mu\nu}:=\delta G_{\mu\nu}+\frac{\alpha}{\kappa}\delta C_{\mu\nu}-\frac{1}{2\kappa}\delta T^{\vartheta}_{\mu\nu}=0
\end{equation}
\\[2pt]

\noindent 2.~{\bf Angular decomposition of the perturbation equations:}
To separate the angular dependence, we expand the metric perturbations $h_{\mu\nu}$ and the pseudoscalar perturbation $\delta\vartheta$ in terms of spherical harmonics. For a slowly rotating background, the equations are typically decomposed using scalar, vector, and tensor spherical harmonics appropriate for the spin-weighted perturbations on a sphere~\cite{Thorne:1980ru,ZERILLI}. For instance, the scalar perturbations ($h_{tt}, h_{tr}, h_{rr}, h_{\theta\theta}+h_{\phi\phi}/\sin^2\theta$)  and pseudoscalar field perturbation ($\delta\vartheta$) are expanded in scalar spherical harmonics $Y_{lm}(\theta, \phi)$. Due to the rotation and dCS effects, harmonics with different $l$ values couple~\cite{Manu}, though for slow rotation, we primarily focus on a given $l$ and its coupling to $l \pm 1, l \pm 2$ induced by rotation and dCS terms. The $m$ quantum number remains a separation constant. \\[2pt]

\noindent 3.~{\bf Simplify the equations by eliminating terms that do not contribute to the QNM frequency at $\mathcal{O}(\chi^2, \alpha^2)$:}
As we are interested in QNM frequency corrections $\delta\omega = \delta\omega_{\chi} + \delta\omega_{\chi^2} + \delta\omega_{\alpha} + \delta\omega_{\alpha^2} + \delta\omega_{\chi\alpha} + ...$, we consistently keep all terms in the perturbation equations that can contribute to the desired order. Terms of $\mathcal{O}(\chi^3)$, $\mathcal{O}(\alpha^3)$, $\mathcal{O}(\chi^3\alpha^2)$, $\mathcal{O}(\chi^2 \alpha^3)$ can be dropped. \\[2pt]

\noindent  4.~{\bf Eliminate the dependent functions by expressing them in terms of independent functions:}
After angular decomposition, we will have a system of coupled ordinary differential equations (ODEs) for various radial functions representing the components of $h_{\mu\nu}$ (e.g., $H_0(r), H_1(r), H_2(r), K(r), h_0(r), h_1(r)$) \cite{Chandrasekhar,Martel:2005ir} and the pseudoscalar perturbation $R(r)$. As we show, many of these functions will be related through algebraic constraint equations arising from the gauge choice or certain combinations of the field equations. By systematically solving these constraints, we express dependent functions in terms of a minimal set of independent functions. This process reduces the dimensionality of the system of differential equations.  \\[2pt]

\noindent 5.~{\bf Obtain the final set of linear coupled differential equations:}
By linearly combining the simplified and reduced equations from the previous steps, we obtain a final system of coupled linear homogeneous ordinary differential equations for the chosen master functions. \\[2pt]

\noindent 6.~{\bf Numerically integrate the final set of linear coupled differential equations to obtain QNM frequencies:}
    With the system of ODEs at hand, QNM frequencies $\omega$ are found by imposing appropriate boundary conditions: purely outgoing waves at spatial infinity ($r_* \to \infty$) and purely ingoing waves at the event horizon ($r_* \to -\infty$).

\subsection{Regge-Wheeler Gauge}

As mentioned above, QNMs are the characteristic oscillation modes through which a BH emits energy when subjected to small perturbations. These modes are intrinsically damped due to the dissipation of energy via GWs. To study these perturbations systematically, we use $\epsilon$ to track the scale of the perturbation, ensuring that the perturbed quantities remain small compared to the background metric. The perturbed metric and pseudoscalar field can be expressed as in Eq.~\eqref{pert_metric}. Since we restrict our analysis to linear perturbations, we retain terms only up to $\mathcal{O}(\epsilon)$ in both the metric and scalar field perturbation equations.  

Although the metric perturbation $h_{\mu\nu}$ is symmetric and possesses ten independent components, due to gauge dedundancy, allows us to impose conditions that reduce the number of independent components to six. A standard choice for this purpose is the \emph{Regge-Wheeler gauge}~\cite{REGGEWHEELER,Nollert}, which simplifies the analysis by eliminating non-physical modes.  These six independent components can be further classified based on their behavior under parity transformations into \emph{odd (axial)} and \emph{even (polar)} parity perturbations~\cite{2018-Berti.etal-GRGa}:
\begin{equation*}  
    h_{\mu\nu} = h_{\mu\nu}^{\text{odd}} + h_{\mu\nu}^{\text{even}}.  
\end{equation*}  
Under a parity transformation $(\theta, \phi) \rightarrow (\pi - \theta, \pi + \phi)$, these perturbations transform as:  
\begin{equation}  
    h_{\mu\nu}^{\text{odd}} \rightarrow (-1)^{l+1} h_{\mu\nu}^{\text{odd}} \quad \text{and} \quad h_{\mu\nu}^{\text{even}} \rightarrow (-1)^{l} h_{\mu\nu}^{\text{even}},  
\end{equation}  
where $l$ is the angular momentum quantum number associated with the spherical harmonic decomposition. Assuming a harmonic time dependence $e^{-i\omega t}$ for all perturbed quantities, we expand the metric perturbations in terms of spherical harmonics. The odd-parity and even-parity perturbations take the form:  
{\small 
\begin{widetext}  
\begin{align}\label{hmunu}  
h_{\mu\nu}^{\text{odd}} = \sum_{l,m} \begin{pmatrix}  
    0 & 0 & h^{lm}_{0}(r) S^{lm}_{\theta}(\theta,\phi) & h^{lm}_{0}(r) S^{lm}_{\phi}(\theta,\phi) \\  
    * & 0 & h^{lm}_{1}(r) S^{lm}_{\theta}(\theta,\phi) & h^{lm}_{1}(r) S^{lm}_{\phi}(\theta,\phi) \\  
    * & * & 0 & 0 \\  
    * & * & * & 0  
\end{pmatrix} e^{-i\omega t},&~
h_{\mu\nu}^{\text{even}} = \sum_{l,m} \begin{pmatrix}  
    H^{lm}_{0}(r) & H^{lm}_{1}(r) & 0 & 0 \\  
    * & H^{lm}_{2}(r) & 0 & 0 \\  
    * & * & r^2 K^{lm}(r) & 0 \\  
    * & * & * & r^2 \sin^2\theta \, K^{lm}(r)  
\end{pmatrix} Y^{lm}(\theta,\phi) e^{-i\omega t}.  
\end{align}  
\end{widetext}  
}
Here, $Y^{lm}(\theta,\phi)$ are the standard spherical harmonics, and $S^{lm}_{\theta}$, $S^{lm}_{\phi}$ are the odd-parity vector spherical harmonics defined as:  
\begin{align}  
    S^{lm}_{\theta}(\theta,\phi) &= -\frac{1}{\sin\theta} \partial_{\phi} Y^{lm}(\theta,\phi), \\  
    S^{lm}_{\phi}(\theta,\phi) &= \sin\theta \, \partial_{\theta} Y^{lm}(\theta,\phi).  
\end{align}  
The symmetric components of the matrices are denoted by $*$.  
Similarly, the perturbation of the pseudoscalar field $\delta\vartheta$ is decomposed as:  
\begin{equation}\label{scalarper}  
\delta\vartheta = \sum_{l,m} \frac{R_{lm}(r)}{r} Y_{lm}(\theta,\phi) e^{-i\omega t}.  
\end{equation}  

With the metric and pseudoscalar field perturbations explicitly defined, we substitute them into the dCS gravity field equations \eqref{eq:dCS-fieldeq}. This yields a set of coupled differential equations governing the perturbations:  
The metric perturbation equations for $h_{\mu\nu}$, derived from the modified Einstein equations. The scalar perturbation equation for $\delta\vartheta$, arising from the dCS scalar field equation.  

\subsection{Perturbation Equations} 

Substituting the perturbed metric \eqref{pert_metric} and pseudoscalar field equations \eqref{scalarper} into the dCS field equations \eqref{eq:dCS-fieldeq} yields a set of eleven coupled linear partial differential equations for the perturbation variables. These equations are valid up to $\mathcal{O}(\epsilon)$, representing linear perturbations around the $\mathcal{O}(\chi^2,\alpha^2)$ slowly rotating dCS BH background. To simplify their structure and facilitate analysis, these \emph{metric perturbation} equations can be classified into three distinct groups based on their transformation properties on the spherical ($\theta-\phi$) surface, analogous to how tensorial quantities are decomposed on a sphere~\cite{Thorne:1980ru}.  \\[2pt]

\noindent{\bf Group-I: Scalar-like perturbation equations:}
These equations correspond to scalar-like components of the perturbed Einstein equations and are of the following form: 
%
\begin{align}\label{group1}
& \delta E_{(i)} =\Big(A^{(i)}_{0,l}+A^{(i)}_{1,l}\cos\theta+A^{(i)}_{2,l}\cos^2\theta\Big)Y_{l} (\theta,\phi) \\ 
& \qquad \quad +\Big(B^{(i)}_{1,l}+B^{(i)}_{2,l}\cos\theta\Big)\sin\theta\partial_{\theta}Y_{l}(\theta,\phi)=0 \, , \nonumber
\end{align}
%
Here, $Y_l(\theta,\phi)$ denotes the spherical harmonics $Y_{lm}(\theta,\phi)$, where the $m$ index is implicitly summed over due to the rotational symmetry of the background, as is typical in such expansions. Specifically, in the above equation $\delta E_{i}$ represents  $\delta E_{tt}$, $\delta E_{tr}$, $\delta E_{rr}$, and $\delta E_{\theta\theta} + \delta E_{\phi\phi}/\sin^2\theta$ for $i=0,1,2,3$ respectively.

The coefficients $A^{(i)}_{n,l}$ and $B^{(i)}_{n,l}$ are functions of the radial coordinate $r$ through $H_0$,$H_1$,$H_2$,$K$,$h_0$,$h_1$,$R$, the QNM frequency $\omega$, the BH mass $M$, the spin parameter $\chi$, and the dCS coupling parameter $\alpha$. Their subscripts $n$ indicate the leading order of $\chi$ at which these terms start appearing.\\[2pt]

\noindent{\bf Group-II: Vector-like perturbation equations:}
This group comprises the vector-like components of the perturbed Einstein equations and are of the following form:
\begin{widetext}
\begin{align}\label{group2}
\delta E_{(j\theta)} &=\Bigg(\sum_{n=0}^2\alpha^{(j)}_{n,l}\cos^n\theta+\Tilde{\alpha}^{(j)}_{2,l}\sin^2\theta\Bigg)\partial_{\theta}Y^{l}-\Bigg(\sum_{n=0}^2\beta^{(j)}_{n,l}\cos^n\theta+\Tilde{\beta}^{(j)}_{2,l}\sin^2\theta\Bigg)\frac{\partial_{\phi}Y^{l}}{\sin\theta}+\Big(\eta^{(j)}_{1,l}+\eta^{(j)}_{2,l}\cos\theta\Big)\sin\theta\,Y_{l}\nonumber\\
&\quad+\Big(\chi^{(j)}_{1,l}+\chi^{(j)}_{2,l}\cos\theta\Big)\sin\theta\,W_l +\Big(\xi^{(j)}_{1,l}+\xi^{(j)}_{2,l}\cos\theta\Big)X_l+\gamma^{(j)}_{2,l}\partial_{\theta}(\sin\theta\,X_l)+\rho^{(j)}_{2,l}\partial_{\theta}(\sin^2\theta W_l)\\
\delta E_{(j\phi)}&=\Bigg(\sum_{n=0}^2\beta^{(j)}_{n,l}\cos^n\theta-\Tilde{\beta}^{(j)}_{2,l}\sin^2\theta\Bigg)\partial_{\theta}Y^{l}+\Bigg(\sum_{n=0}^2\alpha^{(j)}_{n,l}\cos^n\theta-\Tilde{\alpha}^{(j)}_{2,l}\sin^2\theta\Bigg)\frac{\partial_{\phi}Y^{l}}{\sin\theta}
+\Big(\zeta^{(j)}_{1,l}+\zeta^{(j)}_{2,l}\cos\theta\Big)\sin\theta\,Y_{l}\nonumber\\
&\quad+\Big(\chi^{(j)}_{1,l}+\chi^{(j)}_{2,l}\cos\theta\Big)X_{l}
-\Big(\xi^{(j)}_{1,l}+\xi^{(j)}_{2,l}\cos\theta\Big)\sin\theta\,W_{l}+\gamma^{(j)}_{2,l}\partial_{\phi}(X_l)+\rho^{(j)}_{2,l}\partial_{\phi}(\sin\theta\,W_l)\nonumber
\end{align}
\end{widetext}
Here, $W_l$ and $X_l$ denote the angular functions related to the tensor spherical harmonics defined in \eqref{WlmXlm}. The coefficients $\alpha^{(j)}_{n,l}, \tilde{\alpha}^{(j)}_{n,l}, \beta^{(j)}_{n,l}, \tilde{\beta}^{(j)}_{n,l}, \eta^{(j)}_{n,l}, \chi^{(j)}_{n,l}, \xi^{(j)}_{n,l}, \gamma^{(j)}_{n,l}, \rho^{(j)}_{n,l}, \zeta^{(j)}_{n,l}$ are radial functions dependent on the BH and dCS parameters.

These arise from the angular components of the perturbed Einstein equations, specifically mapping to $\delta E_{t\theta}, \delta E_{r\theta}$ (where $j=0,1$ respectively) and their $\phi$-counterparts, $\delta E_{t\phi}, \delta E_{r\phi}$. \\[2pt]

\noindent{\bf Group-III: Tensor-like perturbation equations:}
This final group consists of the tensor-like perturbation equations, as given below: 
\begin{widetext}
\begin{align}\label{group3}
\delta E_{(-)}&=(g_{1,l}+g_{2,l}\cos\theta)\sin\theta\partial_{\theta}Y_{l}
-(f_{1,l}+f_{2,l}\cos\theta)\partial_\phi Y_{l}+\Tilde{h}_{2,l}\sin^2\theta Y_{l} 
+(j_{0,l}+j_{1,l}\cos\theta+j_{2,l}\cos^2\theta)W_{l} \\
& -(k_{0,l}+k_{1,l}\cos\theta+k_{2,l}\cos^2\theta)\frac{X_{l}}{\sin\theta} -s_{2,l}\left(\frac{\partial_{\phi}(\cos\theta\,X_{l})}{\sin\theta}+W_l\right)-t_{2,l}\Big(\partial_{\phi}(\cos\theta\,W_{l})-(l(l+1)-2)\cos\theta\partial_\phi Y_{l}\Big) \nonumber \\
\delta E_{\theta\phi}&=(f_{1,l}+f_{2,l}\cos\theta)\sin\theta\partial_{\theta}Y_{l}+(g_{1,l}+g_{2,l}\cos\theta)\partial_\phi Y_{l}+h_{2,l}\sin^2\theta Y_{l}
+(j_{0,l}+j_{1,l}\cos\theta+j_{2,l}\cos^2\theta)\frac{X_{l}}{\sin\theta} \nonumber\\
&\quad+(k_{0,l}+k_{1,l}\cos\theta+k_{2,l}\cos^2\theta)W_{l} 
+s_{2,l}\Big(\partial_{\theta}(\cos\theta\,X_{l})+(l(l+1)-2)\cos\theta\partial_\phi Y_{l}\Big)+t_{2,l}\Big(\partial_{\theta}(\cos\theta\sin\theta\,W_{l})+W_{l}\Big)\nonumber
\end{align}
\end{widetext}
The coefficients $g,f,h,\tilde{h},j,k,s,t$ are radial functions, similar to those in Group-II. These typically arise from linear combinations of the $\theta\theta$, $\phi\phi$, and $\theta\phi$ components of the perturbed Einstein equations. Specifically, $\delta E_{(-)}$ corresponds to $\delta E_{\theta\theta}-\delta E_{\phi\phi}/\sin^2\theta$.

In addition to these metric perturbation equations, an additional equation is obtained by substituting the \emph{perturbed scalar field} $\delta\vartheta$ into the field equation \eqref{eq:dCS-fieldeq}. This scalar field perturbation equation is itself \emph{scalar-like and naturally falls into the Group-I category} of equations.

The coefficients $A,B,\alpha,\beta,\eta,\chi, \xi,\gamma,\rho,g,f,h,\tilde{h},j,k,s,t$ appearing in the above equations are linear combinations of the radial functions defining the metric perturbations ($H_0,H_1,H_2, K,h_0,h_1$) and the scalar field perturbation ($R$), as well as their derivatives with respect to $r$. The numerical subscripts on these coefficients (e.g., $A^{(i)}_{0,l}$, $A^{(i)}_{1,l}$, $A^{(i)}_{2,l}$) indicate the leading order in the spin parameter $\chi$ at which these terms first appear in the perturbative expansion.

It's important to note that the above equations still contain an intrinsic summation over the angular momentum quantum number $l$ and remain explicitly dependent on the angular coordinates $\theta$ and $\phi$. The next crucial step in solving these perturbation equations is to perform an angular decomposition. This process projects the equations onto a basis of spherical harmonics, effectively removing the angular dependence and transforming the partial differential equations into a set of coupled ordinary differential equations solely dependent on the radial coordinate $r$. This reduction is essential for numerically or analytically solving for the radial functions and subsequently determining the complex quasinormal frequencies $\omega$.

\subsection{Angular decomposition}\label{subsection6.2}

To transform the partial differential equations obtained in the previous subsection into a set of coupled ordinary differential equations, we employ the technique of angular decomposition. This procedure is conceptually similar to that utilized for $\mathcal{O}(\chi)$ perturbations in GR~\cite{Pani}, but with the significant addition of terms arising from the dCS coupling and the second-order spin corrections. Consequently, this necessitates the computation of numerous additional angular integrals beyond those typically encountered in GR. Most of these integrals are systematically cataloged in Ref.~\cite{franchini2023slow}. Furthermore, dCS gravity introduces a few angular integral operators, such as $\bar{\mathcal{X}}, \bar{\mathcal{W}},\mathcal{T},\mathcal{S}, \bar{\mathcal{T}}$, and $\bar{\mathcal{S}}$, which are not present in GR. All relevant angular integral operators, including these dCS-specific ones, are comprehensively listed in Appendix \ref{Appendix_B}.

The angular decomposition is performed by projecting the three groups --- Group I, Group II and Group III --- of perturbation equations onto appropriate angular basis functions. Specifically, we use \emph{spherical harmonics} $Y_{lm}(\theta,\phi)$ to decompose scalar-like (Group-I) equations. For vector-like (Group-II) equations, we employ \emph{vector spherical harmonics}, and finally, for the tensor-like (Group-III) equations, we utilize the specific angular functions $W_{lm}(\theta,\phi)$ and $X_{lm}(\theta,\phi)$ (defined below) which are derived from tensor spherical harmonics~\cite{Thorne:1980ru}.

The angular decomposed equations are given by:
\begin{align}\label{group1_ang_decomp}
\delta E^{I}_{(i+)} & = \int \diff\Omega\,\,\delta E_{(i)}\,Y^{*}_{lm} \nonumber \\ &=\sum_{n=0}^2\Big[\mathcal{C}_{n}A^{(i)}_{n,l}+\mathcal{S}_{n}B^{(i)}_{n,l}\Big] = 0 \\
& \mbox{(Group-I Angular Decomposition)} \nonumber 
\end{align}
where $i=0,1,2,3$ as defined in Eq. \eqref{group1}.
\begin{align}\label{group2_ang_decomp_plus}
& \delta E^{II}_{(j+)} = \int\diff\Omega\,\Bigg(\delta E_{(j\theta)}\,\partial_{\theta}Y^{*}_{lm}+\frac{\delta E_{(j\phi)}\,\partial_{\phi}Y^{*}_{lm}}{\sin\theta}\Bigg)\\
& \qquad~~ =\sum_{n=0}^2\Big[\mathcal{A}_{n}\alpha^{(j)}_{n,l}-\mathcal{B}_{n}\beta^{(j)}_{n,l}\Big]+\Tilde{\mathcal{A}}_{2}\Tilde{\alpha}^{(j)}_{2,l}-\Tilde{\mathcal{B}}_2\Tilde{ \beta^{(j)}_{2,l}}\nonumber\\
&\quad +\sum_{n=1}^{2}\Big[\bar{\mathcal{S}}_n\eta^{(j)}_{n,l}-im\mathcal{C}_{n-1}\zeta^{(j)}_{n,l}+\mathcal{X}_{n-1}\xi^{(j)}_{n,l}+\bar{\mathcal{X}}_{n-1}\chi^{(j)}_{n,l}\Big]\nonumber\\
&\qquad~~+l(l+1)\Big[\bar{\mathcal{X}} \gamma_{2,l}^{(j)}+\bar{\mathcal{W}} \rho_{2,l}^{(j)}\Big] = 0 \nonumber
\end{align}
\begin{align}
& \delta E^{II}_{(j-)} = \int\diff\Omega\,\Bigg(\delta E_{(j\phi)}\,\partial_{\theta}Y^{*}_{lm} - \frac{\delta E_{(j\theta)} \,\partial_{\phi}Y^{*}_{lm}}{\sin\theta}\Bigg)\\
&\qquad~~= \sum_{n=0}^2\Big[\mathcal{A}_{n}\beta^{(j)}_{n,l}+\mathcal{B}_{n}\alpha^{(j)}_{n,l}\Big]-\Tilde{\mathcal{A}}_{2}\Tilde{\beta}^{(j)}_{2,l}-\Tilde{\mathcal{B}}_2\Tilde{\alpha^{(j)}_{2,l}}\nonumber\\
&+\sum_{n=1}^{2}\Big[\bar{\mathcal{S}}_n\zeta^{(j)}_{n,l}+im\mathcal{C}_{n-1}\eta^{(j)}_{n,l}+\mathcal{X}_{n-1}\chi^{(j)}_{n,l}-\bar{\mathcal{X}}_{n-1}\xi^{(j)}_{n,l}\Big] = 0 \nonumber \\
& \quad \mbox{(Group-II Angular Decomposition)} \nonumber 
\end{align}
where $j=0,1$ as defined in Eq.~\eqref{group2}.
\begin{align}\label{group3_ang_decomp_plus}
& \delta E^{III}_{(+)} = \int\diff\Omega\,\Bigg(\delta E_{(-)}\,W^*_{lm}+\frac{\delta E_{(\theta\phi)}\,X^*_{lm}}{\sin\theta}\Bigg) \\ 
&\quad =\sum_{n=0}^1\Big[\mathcal{F}_{n}f_{n+1,l} +\mathcal{G}_{n}g_{n+1,l}\Big] +\sum_{n=0}^2\Big[\mathcal{J}_nj_{n,l}+\mathcal{K}_nk_{n,l}\Big]\nonumber\\
& \quad +\mathcal{W}\tilde{h}_{2,l}+\mathcal{X}h_{2,l} +\mathcal{T}t_{2,l}+\mathcal{S}s_{2,l} = 0 \nonumber \\
& \delta E^{III}_{(-)} = \int\diff\Omega\,\Bigg(\delta E_{(\theta\phi)}\,W^*_{lm} - \frac{\delta E_{(-)} \,X^*_{lm}}{\sin\theta}\Bigg) \\
&=\sum_{n=0}^1\Big[\mathcal{G}_{n}f_{n+1,l} 
-\mathcal{F}_{n}g_{n+1,l}\Big]+\sum_{n=0}^2\Big[\mathcal{J}_n k_{n,l} -\mathcal{K}_nj_{n,l}\Big]\nonumber\\
&\quad+\mathcal{W}h_{2,l}-\mathcal{X}\tilde{h}_{2,l} +\bar{\mathcal{T}}t_{2,l}+\bar{\mathcal{S}}s_{2,l} = 0 \nonumber \\
& \qquad \mbox{(Group-III Angular Decomposition)} \nonumber 
\end{align}

The angular functions $W_{l}$ and $X_{l}$ are derived from the derivatives of spherical harmonics and are given by:
\begin{align}\begin{split}\label{WlmXlm}
W_l&=\partial^2_{\theta}Y_{l}-\cot\theta\,\partial_{\theta}Y_{l}-\frac{1}{\sin^2\theta}\partial_{\phi}^2Y_{l}\\
X_l&=2(\partial_{\theta}\partial_{\phi}(Y_{l})-\cot\theta\partial_{\phi}(Y_{l}))\end{split}\end{align}
The operators 
$\mathcal{C}_n$, $\mathcal{S}_n$, $\mathcal{A}_n$, $\mathcal{B}_n$, $\tilde{\mathcal{A}}_n$, $\tilde{\mathcal{B}}_n$, $\bar{\mathcal{S}}_n$, $\mathcal{X}_{n}$, $\bar{\mathcal{X}}_{n}$, $\bar{\mathcal{X}}$, $\bar{\mathcal{W}}$, $\mathcal{F}_n$, $\mathcal{G}_n$, $\mathcal{J}_n$, $\mathcal{K}_n$, $\mathcal{W}$, $\mathcal{X}$, $\mathcal{T}$, $\mathcal{S}$, $\bar{\mathcal{T}}$, $\bar{\mathcal{S}}$ are the results of specific angular integrations involving products of spherical harmonics and trigonometric functions. Their explicit forms are provided in Appendix \ref{Appendix_B}. Crucially, after this angular decomposition, the equations are purely radial, with no remaining angular dependence.

However, due to the non-spherically symmetric nature of the spinning background metric, particularly at $\mathcal{O}(\chi^2)$, these radial equations exhibit coupling between different multipole moments. Specifically, the equations for a given $l$ mode will be coupled to terms involving $l \pm 1$ and $l \pm 2$ modes. This means that solving for the quasinormal modes requires simultaneously solving a system of coupled radial differential equations for multiple $l$ values.

The general structure of an axial-led radial perturbation equation (where the primary perturbation is odd-parity) would be of the form:
\begin{align}
\begin{split}\label{pert_eqns_axial}
&\mathcal{A}_{lm}+\chi m\bar{\mathcal{A}}_{lm}+\chi^2\hat{\mathcal{A}}_{lm}+\chi^2m^2\bar{\bar{\mathcal{A}}}_{lm}\\
&\quad+\chi\Big(Q_{l-1m}\tilde{\mathcal{P}}_{(l-1)m}+Q_{l+1m}\tilde{\mathcal{P}}_{(l+1)m}\Big)\\
&\quad+\chi^2\Big(Q_{l-1m}Q_{l-2m}\breve{\mathcal{A}}_{(l-2)m}+Q_{l+1m}Q_{l+2m}\breve{\mathcal{A}}_{(l+2)m}\Big)\\
&\quad+m\chi^2\Big(Q_{l-1m}\check{\mathcal{P}}_{(l-1)m}+Q_{l+1m}\check{\mathcal{P}}_{(l+1)m}\Big) = 0
\end{split}
\end{align}
Here, $\mathcal{A}_{lm}$ represents a linear combination of the odd-parity (axial) radial functions ($h_0, h_1, R$) and their derivatives, corresponding to the $(l,m)$ mode. $\mathcal{P}_{lm}$ denotes a linear combination of the even-parity (polar) radial functions ($H_0, H_1, H_2, K$) and their derivatives. The terms containing $Q$ factors illustrate the angular coupling, where $Q_{lm} = \sqrt{(l^2-m^2)/(4l^2-1)}$ are known angular coupling coefficients. The coefficients $\bar{\mathcal{A}}_{lm}, \hat{\mathcal{A}}_{lm}, \bar{\bar{\mathcal{A}}}_{lm}, \tilde{\mathcal{P}}_{(l\pm1)m}, \breve{\mathcal{A}}_{(l\pm2)m}, \check{\mathcal{P}}_{(l\pm1)m}$ encapsulate the complex radial dynamics and inter-mode coupling.

Similarly, a polar-led equation (where the primary perturbation is even-parity) takes the general form:
\begin{align}
\begin{split}\label{pert_eqns_polar}
&\mathcal{P}_{lm}+\chi m\bar{\mathcal{P}}_{lm}+\chi^2\hat{\mathcal{P}}_{lm}+\chi^2m^2\bar{\bar{\mathcal{P}}}_{lm}\\
&\quad+\chi\Big(Q_{l-1m}\tilde{\mathcal{A}}_{(l-1)m}+Q_{l+1m}\tilde{\mathcal{A}}_{(l+1)m}\Big)\\
&\quad+\chi^2\Big(Q_{l-1m}Q_{l-2m}\breve{\mathcal{P}}_{(l-2)m}+Q_{l+1m}Q_{l+2m}\breve{\mathcal{P}}_{(l+2)m}\Big)\\
&\quad+m\chi^2\Big(Q_{l-1m}\check{\mathcal{A}}_{(l-1)m}+Q_{l+1m}\check{\mathcal{A}}_{(l+1)m}\Big) = 0
\end{split}
\end{align}
This explicit coupling between different $l$ modes, up to $l \pm 2$, is a direct consequence of considering second-order spin corrections in the background metric and the intricate nature of the dCS field equations. Solving these coupled radial equations with QNM boundary conditions is the main computational challenge in determining the QNM frequencies of dCS BHs which is the focus of the next section.

\section{QNM frequencies for slowly rotating dCS BHs}

The determination of QNM frequencies for slowly rotating BHs up to first order in $\chi$ often relies on perturbative extensions of spherically symmetric methods. Such approaches have been successfully applied in modified gravity theories, including dynamical Chern-Simons (dCS) gravity~\cite{Manu,wagle}. At $\mathcal{O}(\chi)$, the dominant effect is typically a lifting of the $m$-degeneracy present in the Schwarzschild spectrum, yielding frequencies of the form:  
\[
\omega_{nlm} = \omega_{nl}^{(0)} + m \chi \delta\omega^{(1)} + \mathcal{O}(\chi^2),  
\]  
where $m$ is the azimuthal quantum number, and $\omega_{nl}^{(0)}$ denotes the Schwarzschild QNM frequencies.  

However, as mentioned earlier, extending these calculations to second order in spin, $\mathcal{O}(\chi^2)$, introduces significant complexities that demand a more sophisticated treatment of the perturbation equations. Below, we outline the key challenges and their implications for our further analysis: 
\begin{enumerate}[leftmargin=0.45cm]
\item {\bf Essential $l \leftrightarrow l \pm 1$ Mode Couplings:}  In a spherically symmetric background, perturbations with different angular momentum numbers $l$ decouple. Slow rotation introduces terms proportional to $\chi$ (and angular derivatives like $\chi \sin\theta$ or $\chi \cos\theta$), which couple an $l$-mode primarily to its neighbors, $l \pm 1$ \cite{Kojima,Cutler:1991ab,Pani}.
While these couplings are formally $\mathcal{O}(\chi)$, their contribution to the frequency shift of the primary $l$-mode can effectively enter at $\mathcal{O}(\chi^2)$ when solved perturbatively.  

Crucially, obtaining the QNM frequency accurately to $\mathcal{O}(\chi^2)$ requires retaining these couplings in full --- they cannot be ignored. Their influence, whether through direct contributions or iterative effects in the coupled system, is indispensable for the second-order correction. In dCS gravity, these coupling coefficients further depend on the coupling parameter $\alpha$, intertwining spin and beyond-GR effects.  

\item {\bf Emergence of $l \leftrightarrow l \pm 2$ Couplings at $\mathcal{O}(\chi^2)$:~} 
At second order in spin, new coupling structures arise in the perturbation equations, including direct couplings between $l$-modes and $l \pm 2$-modes~\cite{Kojima,Cutler:1991ab}. These $l \leftrightarrow l\pm2$ couplings are typically absent at $\mathcal{O}(\chi)$~\cite{Manu} but are intrinsic to the second-order rotational effects. They originate from the quadrupole deformation of the spacetime induced by rotation ($M\chi^2$) and other second-order frame-dragging effects. For instance, terms like $\chi^2 P_2(\cos\theta)$ (where $P_2$ is a Legendre polynomial) in the background metric or potentials naturally lead to such couplings when decomposed into spherical harmonics. Within dCS gravity, these $\chi^2$ terms in the perturbation equations will also be modified by $\alpha$, potentially leading to $\alpha^2 \chi^2$ contributions to the coupling potentials. 

\item {\bf Obstacles to decoupling via basis transformations:} 
A common strategy for simplifying coupled systems is to seek a basis transformation that diagonalizes the equations, effectively decoupling the modes. However, for dCS perturbations at $\mathcal{O}(\chi^2,\alpha^2)$, we find that no such transformation can simultaneously eliminate all couplings between $l$, $l \pm 1$, and $l \pm 2$ modes. This reflects the breaking of underlying symmetries that permit decoupling in simpler cases (e.g., spherical symmetry or first-order rotation).  

The persistence of these couplings suggests that mode interactions are intrinsic to $\mathcal{O}(\chi^2)$ perturbations, particularly in modified gravity theories like dCS, where additional fields and couplings further complicate the system. While targeted transformations may simplify subsets of terms, a full decoupling remains unattainable without sacrificing $\mathcal{O}(\chi^2)$ accuracy.  
\end{enumerate}

Given these intrinsic comp lexities, particularly the persistent coupling among different multipoles up to $l\pm2$, an analytical solution for the QNM frequencies at $\mathcal{O}(\chi^2)$ becomes exceedingly difficult. Therefore, we have resorted to \emph{numerical integration} of the system of coupled ordinary differential equations governing the perturbations. As we show in the rest of the section, this approach allows us to directly incorporate all relevant coupling terms ($l \leftrightarrow l\pm1$ and $l \leftrightarrow l\pm2$) and solve the resulting boundary value problem for the complex QNM frequencies $\omega$ to the desired $\mathcal{O}(\chi^2)$ accuracy. Standard numerical techniques, such as the shooting method adapted for coupled systems or direct integration with appropriate boundary conditions at the horizon and infinity~\cite{Pani}, are employed for this purpose. This robust numerical framework is essential for capturing the subtle second-order spin effects on QNMs, which are crucial for precision tests of gravitational theories in the next-generation GW detectors.

\subsection{Boundary conditions}

To determine the QNM frequencies ($\omega = \omega_R + i\omega_I$), we must solve the system of metric perturbation equations (obtained in the previous section) subject to physically appropriate boundary conditions --- perturbations correspond to purely outgoing waves at spatial infinity ($r \to \infty$) and purely ingoing waves at the event horizon ($r \to r_H$)~\cite{Chandrasekhar,Kokkotas:1999bd}. This ensures that no radiation is entering the spacetime from infinity {(purely outward flux of energy-momentum)}, nor is any radiation spuriously emerging from inside the horizon.

{This physical requirement on the energy flux is implemented by imposing specific kinematic conditions on the perturbation fields themselves: they must correspond to purely outgoing waves at spatial infinity ($r \to \infty$) and purely ingoing waves at the event horizon ($r \to r_h$). For a radiative system where the energy flux is quadratic in the field amplitudes --- as is the case for both the gravitational and scalar perturbations in this work --- the condition of purely outgoing waves is the direct mathematical implementation of the condition of a purely outward energy flux. This standard model definition ensures that no radiation is entering the spacetime from infinity, nor is any radiation spuriously emerging from the horizon, effectively representing energy dissipation from the black hole~\cite{Chandrasekhar,Kokkotas:1999bd}.}

In the asymptotic regions near the horizon and at spatial infinity, the coupling terms between different field perturbations and between perturbations with different angular momentum quantum numbers $l$ (i.e., the $l, l\pm1, l\pm2$ couplings discussed earlier) become sub-dominant compared to the kinetic and leading-order potential terms. This simplification allows the system of coupled perturbation equations for the radial functions (denoted generically by $Z_{lm}$, which could represent specific functions like $R_{lm}, h_{0lm}, K_{lm}$) to asymptotically reduce to a set of decoupled wave-like equations. Schematically, they take the form:
\begin{align}
    \begin{split}\label{pert_eqns_tort_enhanced}
        \frac{d^2Z_{lm}}{dr_*^2} + k_H^2 Z_{lm} & 
= \mathcal{O}(r-r_H) \quad \text{as } r \to r_H \\
\frac{d^2Z_{lm}}{dr_*^2} + \omega^2 Z_{lm} & 
= \frac{l(l+1)}{r^2}Z_{lm}+\mathcal{O}\Bigg(\frac{1}{r^3}\Bigg) \quad \text{as } r \to \infty
\end{split}
\end{align}
where the effective potentials (and the coupling terms) vanish sufficiently fast. The parameter $k_H$ is the effective frequency of the wave near the horizon as seen by a co-rotating observer:
\begin{equation}
    k_H = \omega - m\Omega_H
\end{equation}
where $\omega$ is the (complex) QNM frequency, and $\Omega_H$ is the angular velocity of the BH horizon, defined by the limit~\cite{Chandrasekhar}:
\begin{equation}
    \Omega_H = -\lim_{r\to r_H}\frac{g_{t\phi}}{g_{\phi\phi}}
\end{equation}
Using the slowly rotating dCS BH metric components, $k_H$ simplifies to:
\begin{equation}\label{eq:kH_simplified}
    k_H=\omega-\frac{\chi m}{4M}\Bigg(1-\frac{709\alpha^2}{7168\kappa M^4}\Bigg)
\end{equation}
With these definitions, the QNM boundary conditions for each radial function $Z_{lm}$ are:
\begin{align}
    \begin{split}\label{pert_eqn_bcs_enhanced}
        Z_{lm}(r_*) &\sim A_{lm}^{in} e^{-ik_Hr_*} \quad \text{as } r_* \to -\infty \\
        Z_{lm}(r_*) &\sim A_{lm}^{out} e^{+i\omega r_*} \quad \text{as } r_* \to +\infty
    \end{split}
\end{align}
where $A_{lm}^{in}$ and $A_{lm}^{out}$ are complex amplitudes.

The tortoise coordinate $r_*(r)$ is defined by the differential relation $dr_* = dr/F(r)$, where $F(r)$ is chosen such that the set of coupled radial equations take the form Eq. \eqref{pert_eqns_tort_enhanced} close to horizon and at radial infinity~\cite{Pierini:2021jxd,edgbqnm2}. Specifically, $F(r)$ is related to components of the background metric and its asymptotic behavior is crucial:
\begin{itemize}
    \item Near the horizon ($r \to r_H$): $F(r) \propto (r-r_H)$, ensuring $r_* \sim \ln(r-r_H) \to -\infty$.
    \item Far from the BH ($r \to \infty$): $F(r) \to 1$ (for asymptotically flat spacetimes, after appropriate radial scaling if any), ensuring $r_* \sim r \to \infty$.
\end{itemize}
$F(r)$ can be found by requiring that the perturbation equations reduce to Eq. (\ref{pert_eqns_tort_enhanced}), i.e.,
\begin{align}
    \begin{split}
        F(r)&=\Bigg(1-\frac{r_H}{r}\Bigg)\Bigg[1-\chi^2\Bigg(\frac{r_H(r^2+rr_H+r_H^2)}{8r^3}\\
        &\quad+\frac{\alpha^2}{\kappa}\frac{r_H(915r^2+915rr_H-5169r_H^2)}{57344r^3}\Bigg)\Bigg]
    \end{split}
\end{align}
where $r_H$ is given in Eq.~\eqref{eq:horizondCS}. 

{Finally, it is important to note that the axial led \eqref{pert_eqns_axial} and polar led equations \eqref{pert_eqns_polar} are invariant under the transformation $(\chi, m) \to (-\chi, -m)$ with axial perturbations changing their sign and polar perturbations remaining the same.} In other words, the angular decomposed perturbation equations (\ref{pert_eqns_axial}) and (\ref{pert_eqns_polar})) and the boundary conditions (\ref{pert_eqn_bcs_enhanced}) remain invariant under this transformation directly implies that the QNM frequencies $\omega_{nlm}(\chi)$ must satisfy $\omega_{nl(-m)}(-\chi) = \omega_{nlm}(\chi)$. For solutions expanded in powers of $\chi$, this leads to the following general structure for the QNM frequencies:
\[
\omega \simeq \omega_0(n,l) + m\chi\omega_1(n,l) + \chi^2(\omega_{2a}(n,l) + m^2\omega_{2b}(n,l)) 
\]
Here, $\omega_0$ is the QNM frequency for the Schwarzschild BH. The terms $\omega_1$, $\omega_{2a}$, and $\omega_{2b}$ are independent of $m$ and $\chi$, and represent the first-order rotational splitting, the $m$-independent second-order rotational correction, and the $m^2$-dependent second-order rotational correction, respectively. These coefficients will also depend on the dCS parameter $\alpha$. For instance, $\omega_0 = \omega_0^{GR} + \delta\omega_0^{dCS}(\alpha)$, and similarly for $\omega_1, \omega_{2a}, \omega_{2b}$. 

\subsection{Numerical Integration}

As mentioned earlier, the core challenge in determining QNM frequencies at $\mathcal{O}(\chi^2)$ lies in managing the complex system of coupled perturbation equations. Our numerical approach is guided by a careful analysis of which terms contribute to the QNM frequencies at the desired order of accuracy.

It is known that for slowly rotating BHs, while $\mathcal{O}(\chi)$ terms in the perturbation equations couple an $l$-indexed mode to $l\pm1$ modes, these couplings affect the QNM frequency of the primary $l$-mode starting at $\mathcal{O}(\chi^2)$ if treated perturbatively, or are essential components of an $\mathcal{O}(\chi)$ calculation if the system is solved directly. For $\mathcal{O}(\chi)$ frequency corrections, a simpler analysis often suffices~\cite{Manu,wagle,Pani}.

When targeting QNM frequencies accurate to $\mathcal{O}(\chi^2)$, we must account for higher-order effects, we need to follow a different procedure to simplify the system \cite{edgbqnm2}:
\begin{enumerate}
\item Assume the system is initially excited by a purely axial perturbation with a specific harmonic index $l$. In the non-rotating limit ($\chi=0$), all polar parity perturbations ($Z^{l'm}_{\text{pol}}$) are zero, and axial parity perturbations ($Z^{l'm}_{\text{ax}}$) are zero for $l' \neq l$.
\item At $\mathcal{O}(\chi)$, rotation induces couplings that excite polar parity functions with harmonic indices $l\pm1$. Thus, the amplitudes of these modes are $Z^{l\pm 1m}_{\text{pol}} = \mathcal{O}(\chi)$.
\item At $\mathcal{O}(\chi^2)$, rotation can also induce couplings between the primary axial $l$-mode and axial $l\pm2$ modes. Consequently, the amplitudes of these modes are $Z^{l\pm 2m}_{\text{ax}} = \mathcal{O}(\chi^2)$.
\item The crucial point for simplification is that the back-reaction of these $\mathcal{O}(\chi^2)$-amplitude $Z^{l\pm 2m}_{\text{ax}}$ modes onto the QNM frequency of the primary $Z^{lm}_{\text{ax}}$ mode will be of $\mathcal{O}(\chi^4)$ (since the coupling term itself is $\mathcal{O}(\chi^2)$). Therefore, for determining the QNM frequency of $Z^{lm}_{\text{ax}}$ up to $\mathcal{O}(\chi^2)$, the explicit inclusion of $Z^{l\pm 2m}_{\text{ax}}$ modes and the axial $l \leftrightarrow l\pm2$ coupling terms is not necessary.
\end{enumerate}

This systematic ordering allows us to truncate the hierarchy of coupled equations. The final set of perturbation equations, retaining only those terms that contribute to the QNM frequencies of the primary axial $l$-mode up to $\mathcal{O}(\chi^2)$, involves the primary axial $l$-mode and the polar $l\pm1$ modes:
\begin{align}
\begin{split} \label{final_pert_eqns}
    &\mathcal{A}_{lm}+\chi m\bar{\mathcal{A}}_{lm}+\chi^2\hat{\mathcal{A}}_{lm}+\chi^2\bar{\bar{\mathcal{A}}}_{lm}\\
    &+\chi(Q_{lm}\tilde{\mathcal{P}}_{l-1m}+Q_{l+1m}\tilde{\mathcal{P}}_{l+1m})=0\\
    &\mathcal{P}_{l+1m}+\chi m\bar{\mathcal{P}}_{l+1m}+\chi Q_{l+1m}\tilde{\mathcal{A}}_{lm}\\
    &+\chi^2m
Q_{l+1m}\check{\mathcal{A}}_{lm}=0\\
&\mathcal{P}_{l-1m}+\chi m\bar{\mathcal{P}}_{l-1m}+\chi Q_{lm}\tilde{\mathcal{A}}_{lm}\\
&+\chi^2 m
Q_{lm}\check{\mathcal{A}}_{lm}=0
\end{split}
\end{align}
Here, $\mathcal{A}_{lm}$, $\mathcal{P}_{l\pm1m}$ represent the principal differential operators for the axial $l$-mode and polar $l\pm1$ modes, respectively. The terms with bars, hats, tildes, and checks denote various $\mathcal{O}(\chi)$ or $\mathcal{O}(\chi^2)$ corrections to these operators or coupling terms, with $Q$ being coupling coefficients. In this work, we focus on obtaining QNMs of the axial sector, as preliminary investigations and existing literature suggest that for some modified gravity theories, the dominant deviations from QNM frequencies for GR, particularly those introduced by parity-violating terms like in dCS gravity, are more prominently captured in the axial sector due to the parity violation in dCS~\cite{Molina:2010fb}. 

\subsubsection{Numerical method and boundary conditions}

As mentioned earlier, obtaining QNMs is an eigenvalue problem for the complex frequency $\omega$, subject to the boundary conditions of purely ingoing waves at the horizon ($r_* \to -\infty$) and purely outgoing waves at infinity ($r_* \to +\infty$). We numerically integrate the system of equations  using Runge-Kutta-Fehlberg method. The integration is performed from a numerical horizon $r^{\rm num}_{H} = r_H + \delta$ (where $r_H$ is the actual horizon radius and $\delta$ is a small positive number) and from a numerical infinity $r^{num}_{\infty}>>r_H$. For instance, we choose $\delta =0.002M$ and $r^{num}_{\infty}=40M$.

At these numerical boundaries, asymptotic solutions are imposed. These solutions are derived from the behavior of equations  as $r \to r_H$ and $r \to \infty$. After transforming to the tortoise coordinate $r_*$, the asymptotic solutions are plane waves:
\begin{itemize}
\item At the horizon ($r_* \to -\infty$): $Z(r_*) \sim C_{in} e^{-ik_H r_*}$
\item At infinity ($r_* \to +\infty$): $Z(r_*) \sim C_{out} e^{+i\omega r_*}$
\end{itemize}
These forms can be expanded as series solutions in terms of $(r-r_H)$ near the horizon, or $1/r$ at infinity, to provide initial values for the numerical integration in the $r$-coordinate. Specifically, we consider all fields to be ingoing at the horizon,
\begin{equation}
    \sim \sum_j a_j(r-r_H)^je^{-i(\omega t+k_Hr_*)}
\end{equation}
and outgoing at infinity,
\begin{equation}
    \sim\sum_j b_jr^{-j}e^{-i\omega(t-r_*)}
\end{equation}

The system of second-order differential equations in Eqs.~(\ref{final_pert_eqns})  is first reduced to a larger system of first-order differential equations. This involves algebraic manipulations to eliminate dependent functions and their higher derivatives, expressing them in terms of a minimal set of independent functions and their first derivatives. For the axial $l$-mode coupled to polar $l\pm1$ modes, this results in a system of 8 independent first-order ODEs if, for example, the primary axial $l$-mode is described by two first-order functions ($h_0^{lm}, h_1^{lm}$), the scalar mode is given by two functions $R_{lm}$ and $\partial_rR_{lm}$ and each polar $l\pm1$ sector ($H_1^{l\pm1,m}, K^{l\pm1,m}$) effectively contributes two first-order functions.
This system can be written in matrix form:
\begin{equation}
\label{MatrixformDE}
  \frac{d}{dr}\mathbf{\Psi}^l +\hat{A}_l\mathbf{\Psi}^l=0 
\end{equation}
where $\hat{A}_l$ is an 8 dimensional square matrix and $\mathbf{\Psi}^{lm}$ is an 8-component vector of the independent radial functions:
\begin{equation}
\mathbf{\Psi}^{lm}=\begin{pmatrix}    
h_{0}^{lm}\\
h_1^{lm}\\
R^{lm}\\ 
\partial_rR^{lm}\\ 
H_1^{l+1m}\\ K^{l+1m}\\ 
H_1^{l-1m}\\ K^{l-1m}
\end{pmatrix}
\end{equation}

To find the QNM frequencies $\omega$, we perform a Runge–Kutta–Fehlberg integration method. We generate four linearly independent solutions satisfying the ingoing wave boundary condition at $r^{num}_{H}$ and integrate them outwards to a common matching point, say $r_M$. Similarly, we generate four linearly independent solutions satisfying the outgoing wave boundary condition at $r^{num}_{\infty}$ and integrate them inwards to $r_M$, we have two sets of four solution vectors. For a QNM frequency, these two sets must be linearly dependent, meaning a non-trivial linear combination of the solutions from one side and the other side would be zero. This is equivalent to requiring that the determinant of an $8 \times 8$ matrix, whose columns (or rows) are formed by these eight solution vectors evaluated at $r_M$, must vanish: 
\[\det[\mathbf{\Psi}_1^{H}, ..., \mathbf{\Psi}_4^{H}, \mathbf{\Psi}_1^{\infty}, ..., \mathbf{\Psi}_4^{\infty}](r_M) = 0.
\]
We employ a numerical root-finding algorithm, such as the secant method, in the complex $\omega$ plane to find frequencies that satisfy this determinant condition.

For the specific case of $l=2$ (corresponding to the dominant quadrupolar gravitational radiation), if the $l-1=1$ polar modes are absent due to the gauge choice, the system can simplify. If $H^{l-1,m}_1=0$ and $K^{l-1,m}=0$ for $l=2$, then the system reduces to the axial $l=2$ mode coupled only to the polar $l+1=3$ mode. In that case, we will have only six differential equations which should be numerically integrated with three independent boundary conditions at horizon and three independent boundary conditions at infinity. We obtain a total of six vectors with each vector's length being six and thereby obtaining a $6\times6$ matrix. We use the root-finding algorithm to find the root of the determinant of the $6\times6$ matrix.

\section{Results}

In this section, we present our numerical results for the QNM frequencies, focusing on the \emph{axial gravitational perturbations} for the fundamental $n=0$, $l=m=2$ mode. This focus is motivated by the nature of dCS gravity; at zeroth order in spin ($\chi=0$), leading-order dCS corrections to QNM frequencies (proportional to $\alpha^2$) manifest in the axial sector, while the polar sector typically remains uncorrected at this specific order \cite{Molina:2010fb}. This makes the axial modes prime candidates for observing dCS signatures.

\subsection{Numerical integration framework and accuracy assessment}

First, to validate the numerical integration framework and assess the accuracy, we computed QNM frequencies in limiting cases:
\begin{enumerate}
\item {\bf Schwarzschild Limit:} Setting $\alpha=0$ and $\chi=0$, our code reproduced the well-established QNM frequencies for a Schwarzschild BH with excellent agreement with previous results, such as those in \cite{Kokkotas:1999bd,blackholesandbranesqnms}.

\item {\bf Slow rotation (Kerr) Limit:} To assess the accuracy of our second-order-in-spin computation, we set $\alpha=0$ and calculated QNM frequencies for a Kerr BH. These were then compared against highly accurate ``exact" QNM frequencies obtained using Leaver's continued fraction method~\cite{Leaver:1985ax}.
\end{enumerate}

\begin{figure}[htbp]
  \centering
  \fbox{\includegraphics{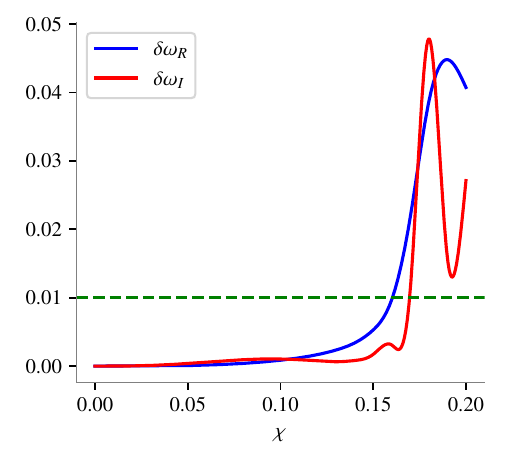}}
  \caption{The relative error between QNM frequencies obtained from our second-order-in-spin numerical integration ($\omega_{\text{approx}}$) and those from Leaver's method for Kerr BHs ($\omega_{\text{exact}}$), for the $n=0, l=m=2$ mode, as a function of the spin parameter $\chi$. Errors for both real ($\delta\omega_R$) and imaginary ($\delta\omega_I$) parts are shown.}
  \label{fig1}
\end{figure}

The relative errors are defined as:
\begin{align}
\delta\omega_R = 1 - \frac{\text{Re}(\omega_{\text{approx}})}{\text{Re}(\omega_{\text{exact}})};
    \delta\omega_I = 1 - \frac{\text{Im}(\omega_{\text{approx}})}{\text{Im}(\omega_{\text{exact}})}
\end{align}
As depicted in \hyperref[fig1]{Fig. 1}, the relative errors for both the real and imaginary parts of the QNM frequency remain below $1\%$ (i.e., $|\delta\omega| < 0.01$) for spin parameter $\chi \leq 0.15$. This demonstrates a significant improvement in the range of validity compared to first-order spin calculations, which, as noted in \cite{wagle}, typically maintain similar accuracy only up to $\chi \leq 0.0375$. As mentioned earlier, the recent work by Chung et al \cite{chung2025quasinormalmodefrequenciesgravitational} have obtained axial and polar QNMs accurate up to $\chi\leq0.75$.
Beyond $\chi \approx 0.15$, we observe that the errors in our second-order approximation begin to grow and exhibit fluctuations, rendering the QNM frequency values less reliable for larger spins. Consequently, for subsequent analyses involving dCS effects, we restrict our study to the domain $\chi \leq 0.15$.

\begin{figure*}
  \centering
  \begin{minipage}[b]{0.45\textwidth}
    \centering
    \fbox{\includegraphics[width=\linewidth]{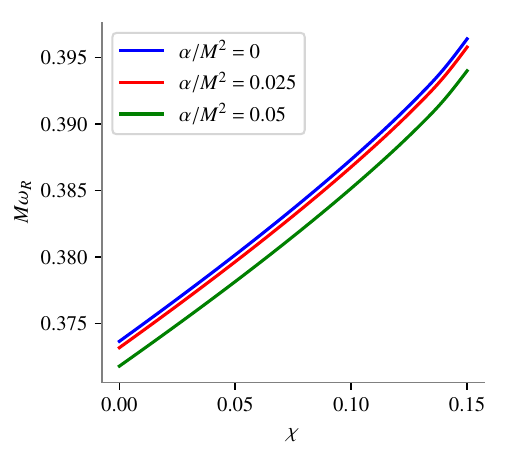}}
  \end{minipage}
  \hfill
  \begin{minipage}[b]{0.45\textwidth}
    \centering
    \fbox{\includegraphics[width=\linewidth]{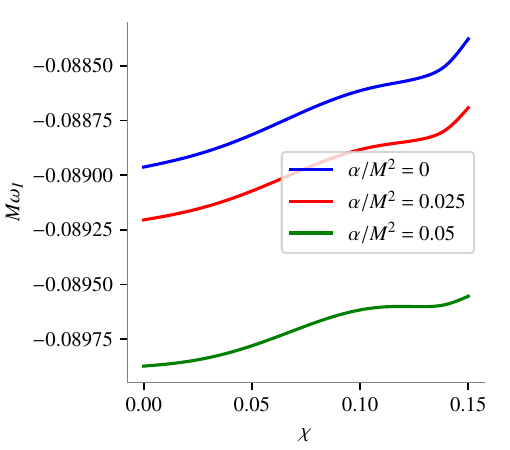}
}  \end{minipage}
  \caption{Real and Imaginary parts of the $n=0$, $l=m=2$ QNM frequency of the gravitational axial mode plotted against the spin parameter $\chi$ for various values of $\alpha/M^2$.}
  \label{fig2}
\end{figure*}
  \begin{figure*}[h!tbp]
  \centering
  \begin{minipage}[b]{0.45\textwidth}
    \centering
    \fbox{\includegraphics[width=\linewidth]{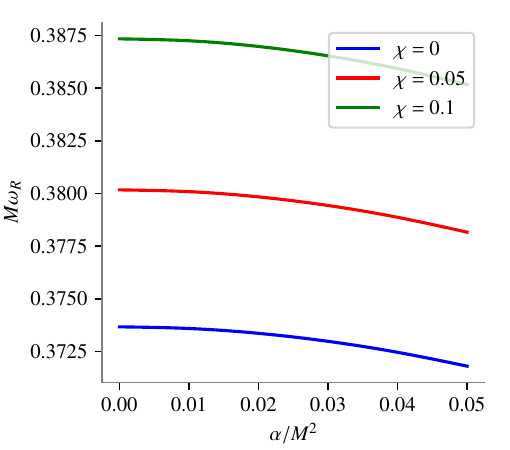}}
  \end{minipage}
  \hfill
  \begin{minipage}[b]{0.45\textwidth}
    \centering
    \fbox{\includegraphics[width=\linewidth]{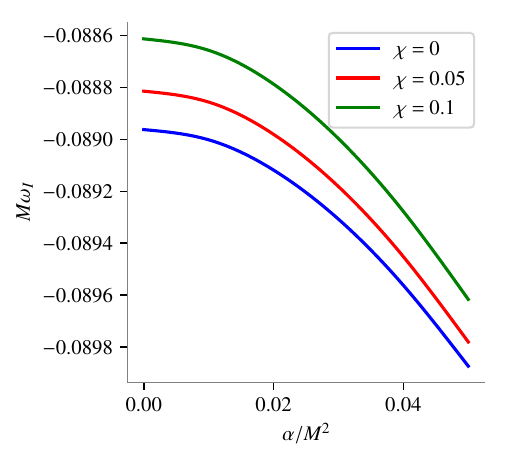}
}  \end{minipage}
  \caption{Real and Imaginary parts of the $n=0$, $l=m=2$ QNM frequency of the gravitational axial mode plotted against $\alpha/M^2$ for various values of the spin parameter $\chi$.}
  \label{fig3}
\end{figure*}

\subsection{dCS QNM Frequencies ($\chi \neq 0, \alpha \neq 0$)}

We now turn to the QNM frequencies in dCS gravity for non-zero spin $(\chi)$ and dCS coupling ($\alpha$). All frequencies are for the $n=0, l=m=2$ axial mode.

In Fig.~\eqref{fig2}, we plot $M\text{Re}(\omega)$ and $M\text{Im}(\omega)$ against the spin parameter $\chi$ for fixed values of the dimensionless dCS coupling $\alpha/M^2 \in \{0, 0.025, 0.05\}$. From the left figure we see that the real part of the QNM frequency, $M\text{Re}(\omega)$, increases with increasing spin for all considered values of $\alpha/M^2$. {From the right figure we see that the imaginary part, $M|\text{Im}(\omega)|$ (which dictates the damping time, $\tau = 1/|\text{Im}(\omega)|$), decreases (damping time increases) with $\chi$.} This trend is qualitatively similar to that observed in GR, where rotation generally increases the oscillation frequency and the damping time of prograde modes ($m>0$)~\cite{Leaver:1985ax}.

To further understand the behavior, in Fig.~\eqref{fig3}, we plot $M\text{Re}(\omega)$ and $M\text{Im}(\omega)$ against $\alpha/M^2$ for \emph{fixed spin parameters} $\chi \in \{0, 0.05, 0.1\}$. From the figure we see that for all spin values considered, $M\text{Re}(\omega)$ is observed to decrease and $M|\text{Im}(\omega)|$ is observed to increase as the dCS coupling $\alpha/M^2$ increases. This implies that the dCS modification, in the regime explored, tends to lower the oscillation frequency and reduce the damping time (increase the damping rate) of the $n=0, l=m=2$ axial mode. This work suggests that modified gravity theories, such as dCS, could lead to \emph{shorter damping times compared to GR}, a distinct signature with potential observability in next-generation gravitational wave detectors.

Our numerically obtained QNM frequencies, particularly their trends, show good qualitative agreement with recent computations using a modified Teukolsky formalism, which were carried out up to first order in spin and second order in the dCS coupling constant \cite{li2025perturbationsspinningblackholes}. It is important to note that while our perturbation equations are constructed to be accurate up to $\mathcal{O}(\chi^2)$ and $\mathcal{O}(\alpha^2)$ in their coefficients, the numerical integration solves these equations without further approximation in the parameters $\chi$ and $\alpha$ (within their range of validity). The resulting QNM frequencies $\omega(\chi, \alpha)$ are thus non-trivial functions, and their graphical representation in the plots naturally exhibits behavior more complex than simple quadratic polynomials.

\subsubsection{Analytic Fitting Formula}
To provide a convenient analytical representation of our results for small $\chi$ and $\alpha/M^2$, we performed a numerical fit to the QNM frequencies obtained for $\chi \in \{0, 0.01, 0.02\}$ and $\alpha/M^2 \in \{0, 0.01, 0.02\}$. For the $n=0, l=m=2$ axial QNM frequency, we obtain the following expression:
\begin{align}
\label{eq:anal-fitting}
\begin{split}
     M\omega&=\omega_{00}+\frac{\alpha^2}{M^4}\omega_{02}+\chi\Bigg(\omega_{10}+\frac{\alpha^2}{M^4}\omega_{12}\Bigg)\\
     &\quad+\chi^2\Bigg(\omega_{20}+\frac{\alpha^2}{M^4}\omega_{22}\Bigg)
\end{split}
\end{align}
where the coefficients are:
\begin{align}
\begin{split}
    \omega_{00} &= 0.37367 - 0.088962i \\
    \omega_{02} &= -0.769 - 0.389i \\
    \omega_{10} &= 0.126 + 0.00192i \\
    \omega_{12} &= -1.13 - 0.565i \\
    \omega_{20} &= 0.078 + 0.016i \\
    \omega_{22} &= -1.7 + 0.66i
\end{split}
\end{align}
The GR coefficients $\omega_{00}$ (Schwarzschild), $\omega_{10}$ (1st order spin correction for $m=2$), and $\omega_{20}$ (2nd order spin correction for $m=2$) are consistent with the literature~\cite{Leaver:1985ax,Manu}. The corresponding pure GR QNMs for a Kerr BH obtained from $\omega_{00}$, $\omega_{10}$ and $\omega_{20}$ are within $1\%$ error for spin values as high as $0.4$~\cite{Leaver:1985ax}. The dCS correction terms $\omega_{02}$, $\omega_{12}$, and $\omega_{22}$ represent new results from this work for the combined effects of dCS coupling and spin up to second order.

We compared these dCS-induced coefficients with those available in Ref. \cite{chung2025quasinormalmodefrequenciesgravitational}. The values for $\omega_{02}$, $\omega_{12}$, and $\omega_{22}$ appear to agree with their findings up to an overall normalization factor ($\pi$), which likely arises from different definitions of the dCS coupling constant $\alpha$ or the Pontryagin density in the respective Lagrangians~\cite{Lagrangian}.  After accounting for a plausible normalization factor, our real parts of the QNM frequencies align reasonably well. However, we note that the imaginary part of our $\omega_{22}$ coefficient ($+0.66i$) appears to differ more substantially from the corresponding value in Ref.~\cite{chung2025quasinormalmodefrequenciesgravitational} than would be explained by a simple overall normalization consistent with other terms. This discrepancy in $\text{Im}(\omega_{22})$ warrants further investigation. Potential sources could include differences in higher-order terms implicitly captured or neglected in the respective theoretical formalisms (e.g. terms beyond $\mathcal{O}(\alpha^2\chi^2)$ in the equations vs. terms in a resummed/alternative approach), the sensitivity of this specific coefficient to the numerical fitting procedure or the range of parameters used, subtle differences in the definition or implementation of the dCS action, or variations in how the coupled system of equations is treated at $\mathcal{O}(\chi^2)$.

The determination of these coefficients, particularly $\omega_{22}$, is a key outcome, as it quantifies the coupled second-order effects of spin and dCS modification on BH ringdown, offering a more refined template for future gravitational wave observations aiming to constrain dCS gravity.

\section{Conclusions and Discussions}

In this work, we have undertaken a comprehensive investigation into the QNMs of slowly rotating BHs within the framework of dCS gravity, extending calculations up to second order in the dimensionless spin parameter $\chi$ and second order in the dCS coupling constant $\alpha$. Recognizing the limitations of traditional Newman-Penrose techniques for constructing tetrads at the second order in $\chi$, we adopted a metric perturbation approach. This method, while algebraically intensive, allowed us to systematically account for the complex couplings between different angular momentum modes ($l, l\pm1, l\pm2$) that become critical at $\mathcal{O}(\chi^2)$.

Our methodology involved deriving the perturbation equations in the Regge-Wheeler gauge, performing an angular decomposition, and then carefully simplifying the system by identifying and retaining only those terms contributing to the QNM frequencies at the desired $\mathcal{O}(\chi^2, \alpha^2)$ accuracy. This led to a final set of coupled linear ordinary differential equations for the primary axial $l$-mode and the polar $l\pm1$ modes, which we solved numerically using a robust Runge-Kutta-Fehlberg method with appropriate ingoing and outgoing wave boundary conditions.
The key results of this study are:
\begin{enumerate}[leftmargin=0.45cm]
\item {\bf Numerical computation up to second order in $\chi$}: We have successfully computed the QNM frequencies for the fundamental $n=0, l=m=2$ axial mode of slowly rotating dCS BHs. Our numerical code was validated against known results for Schwarzschild BHs and, for the Kerr limit ($\alpha=0$) in GR. We also demonstrated accuracy better than $1\%$ for spin parameters $\chi \leq 0.15$ when compared to highly accurate continued fraction methods. This establishes the reliability of our second-order formalism within this spin range.

\item {\bf dCS effects on QNM frequencies:} We presented  numerical results illustrating the dependence of QNM frequencies on both the spin $\chi$ and the dCS coupling $\alpha/M^2$. We observed that both the real and imaginary parts of the QNM frequency increase with spin, similar to GR. 

Our results implies that the dCS modification, in the regime explored, tends to lower the oscillation frequency and \emph{reduce the damping time (increase the damping rate)} of the fundamental ($n=0$), $l=m=2$ axial mode . This finding for the axial mode is particularly noteworthy, as while earlier analytical calculations by Srivastava et al.~\cite{Manu} reported that dCS corrections make the imaginary part of the \emph{polar QNM} less negative (thereby decreasing the decay rate and increasing the damping time), \emph{our analysis for the axial QNM reveals the opposite trend, showing a reduced damping time.} This distinct behavior for different parity modes highlights the rich and complex phenomenology introduced by dCS gravity, with potential observable signatures in next-generation GW detectors.

\item {\bf Analytical Fitting Formula:} We derived an analytical fitting formula \eqref{eq:anal-fitting} for the $n=0$, $l=m=2$ axial QNM frequency, explicitly providing the coefficients $\omega_{00}, \omega_{02}, \omega_{10}, \omega_{12}, \omega_{20},$ and $\omega_{22}$ that capture the zeroth, first, and second-order spin effects, each including zeroth and second-order dCS corrections. These coefficients quantify the intricate interplay between BH spin and dCS modifications. The coefficients $\omega_{00},\omega_{10}$ and $\omega_{20}$ give accurate QNM frequencies of Kerr BH (in the limit $\alpha=0$) for spin values $\chi\leq0.4$.

Our results for the GR limit and the dCS corrections show broad agreement with existing literature, including recent computations using modified Teukolsky formalisms up to first order in spin. We also compared our fitted dCS coefficients with other works, finding agreement for several terms up to normalization factors. However, a discrepancy in the imaginary part of the $\alpha^2\chi^2$ coupling coefficient ($\text{Im}(\omega_{22})$) was noted, highlighting an area for further investigation and cross-validation within the community. Our results  suggests that modified gravity theories, such as dCS, could lead to shorter damping times compared to GR, a distinct signature with potential observability in next-generation gravitational wave detectors.
\end{enumerate}

Our computation of axial QNM frequencies, accurate to second order in both spin and dCS coupling, provides more precise theoretical templates. These templates are crucial for probing potential deviations from GR using GW observations of BH ringdowns. While this analysis could, in principle, be extended to polar modes, we did not pursue in this work. This is because, unlike the axial mode, the polar mode does not exhibit a leading-order correction in spin from dCS effects. Consequently, the dCS correction to the polar QNM frequencies in GR is significantly smaller, making its computational evaluation less pressing for current observational capabilities.

While our current second-order expansion shows high accuracy for $\chi \leq 0.4$, future work is needed to extend this reach to higher spins, potentially through resummation techniques tailored for dCS gravity or by developing fully numerical solutions for rotating dCS BH perturbations. 
In the recent work by Chung et al \cite{chung2025quasinormalmodefrequenciesgravitational}, they have obtained QNMs accurate up to dimensionless spin parameter $\chi\leq0.75$ using the Metric pErTurbations wIth speCtral methodS (METRICS) formalism. However, QNMs of extremal BHs are yet to be probed in dCS theory. Further work could also involve investigating higher overtones, and performing detailed Fisher information analyses to assess the detectability of these second-order effects with current and future gravitational wave observatories.

The recent GW event GW230529~\cite{gw230529}, a low-mass binary BH merger, provides a compelling new avenue to explore potential deviations from GR. Follow-up analyses of this event have estimated the component masses to be relatively small, at $3.6^{+0.8}_{-1.1} \, M_\odot$ and $1.4^{+0.6}_{-0.3} \, M_\odot$. If the resulting remnant is indeed a BH, the QNMs emitted during its ringdown phase offer a unique window into the spacetime geometry and the fundamental theory of gravity. In Ref.~\cite{Manu}, the authors demonstrated that even at linear order in the rotation parameter, dynamical Chern-Simons corrections to the QNMs of a $\sim 15 \, M_\odot$ BH could be significant enough to be potentially detectable. Given the lower mass scale of the final BH formed in GW230529, it is plausible that the dCS corrections to its QNMs could be even more pronounced. This expectation arises from the fact that the magnitude of certain modified gravity effects can scale inversely with the BH mass (e.g., if the dCS coupling $\alpha$ has dimensions of length-squared, then dimensionless quantities often involve $\alpha/M^2$). Consequently, the lower mass of the GW230529 remnant suggests that the subtle deviations from GR predicted by dCS gravity, particularly those encoded in the second-order spin corrections to the BH metric and subsequently in the QNM spectrum, might manifest more distinctly in this mass range. Therefore, investigating the second-order spin corrections to QNMs in dCS gravity becomes particularly relevant and highly motivated in the context of low-mass BH merger events like GW230529, offering a potentially enhanced sensitivity to the effects of this modified gravity theory. Our current work provides the necessary theoretical framework for such future investigations.

\begin{acknowledgments}
This work is part of the Undergraduate project of TA.
The work is supported by SERB-CRG grant.
\end{acknowledgments}

\appendix
\section{Axisymmetric Metric}\label{Appendix_A}
The Kerr metric is given by,
\begin{widetext}
\begin{equation}ds_{K}^2=-\Bigg(1-\frac{2Mr}{\Sigma}\Bigg)dt^2-\frac{4Mar\sin^2\theta}{\Sigma}dtd\phi+\frac{\Sigma}{\Delta}dr^2+\Sigma d\theta^2+\Bigg(r^2+a^2+\frac{2Ma^2r\sin^2\theta}{\Sigma}\Bigg)\sin^2\theta d\phi^2
\end{equation}
\end{widetext}
where, $\Delta=r^2-2Mr+\chi^2M^2\quad\text{and}\quad\Sigma=r^2+\chi^2M^2\cos^2\theta$. This exact Kerr line element can trivially be Taylor expanded up to $\mathcal{O}(\chi^2)$ to obtain the slowly rotating version. 
\begin{widetext}
\begin{align}\label{background_metric}
\begin{split}
ds_{\text{dCS2}}^2&=g^{\text{CS}}_{tt}dt^2+g^{\text{CS}}_{rr} dr^2+g^{\text{CS}}_{\theta\theta}d\theta^2+ g^{\text{CS}}_{\phi\phi}d\phi^2\\
  g^{\text{CS}}_{tt}&=\frac{\alpha^2}{\beta \kappa M^4}\frac{M^3}{r^3}\Bigg[\frac{201}{1792}\Bigg(1+\frac{M}{r}+\frac{4474M^2}{4221r^2}-\frac{2060M^3}{469r^3}+\frac{1500M^4}{469r^4}-\frac{2140M^5}{201r^5}+\frac{9256M^6}{201r^6}-\frac{5376M^7}{67r^7}\Bigg)\times\\
  &\quad(3\cos^2\theta-1)-\frac{5M^2}{384r^2}\Bigg(1+\frac{100M}{r}+\frac{194M^2}{r^2}+\frac{2220M^3}{7r^3}-\frac{1512M^4}{5r^4}\Bigg)\Bigg]\\
  g^{\text{CS}}_{rr}&=\frac{\alpha^2}{\beta \kappa M^4}\frac{M^3}{r^3f_0(r)^2}\Bigg[\frac{201f_0(r)}{1792}\Bigg(1+\frac{1459M}{603r}+\frac{20000M^2}{4221r^2}+\frac{51580M^3}{1407r^3}-\frac{7580M^4}{201r^4}-\frac{22492M^5}{201r^5}-\frac{40320M^6}{67r^6}\Bigg)\times\\
  &\quad(3\cos^2\theta-1)-\frac{25M}{384r}\Bigg(1+\frac{3M}{r}+\frac{322M^2}{5r^2}+\frac{198M^3}{5r^3}+\frac{6276M^4}{175r^4}-\frac{17496M^5}{25r^5}\Bigg)\Bigg]\\
  g^{\text{CS}}_{\theta\theta}&=\frac{201}{1792}\frac{\alpha^2}{\beta \kappa M^4}M^2\frac{M}{r}\Bigg(1+\frac{1420M}{603r}+\frac{18908M^2}{4221r^2}+\frac{1480M^3}{603r^3}+\frac{22640M^4}{1407r^4}+\frac{3848M^5}{201r^5}+\frac{5376M^6}{67r^6}\Bigg)(3\cos^2\theta-1)\\
  g^{\text{CS}}_{\phi\phi}&=\sin^2\theta g^{\text{CS}}_{\theta\theta}
\end{split}
\end{align}
\end{widetext}
\section{Angular Integral formulae}\label{Appendix_B}

As mentioned in Sec.~\eqref{subsection6.2}, the operators $\mathcal{C}_n$, $\mathcal{S}_n$, $\mathcal{A}_n$, $\mathcal{B}_n$, $\tilde{\mathcal{A}}_n$, $\tilde{\mathcal{B}}_n$, $\bar{\mathcal{S}}_n$, $\mathcal{X}_{n}$, $\bar{\mathcal{X}}_{n}$, $\bar{\mathcal{X}}$, $\bar{\mathcal{W}}$, $\mathcal{F}_n$, $\mathcal{G}_n$, $\mathcal{J}_n$, $\mathcal{K}_n$, $\mathcal{W}$, $\mathcal{X}$, $\mathcal{T}$, $\mathcal{S}$, $\bar{\mathcal{T}}$, $\bar{\mathcal{S}}$ are the results of specific angular integrations involving products of spherical harmonics and trigonometric functions. As given below, all these operators are defined in terms of $Q_l$ which is given by:
\begin{equation}
Q_{l}=\sqrt{\frac{l^2-m^2}{4l^2-1}} \, ,
\end{equation}
and $f_l$'s are the functions of $l$. 
\subsection{Group 1}

\begin{flushleft}
\begin{equation*}\mathcal{C}_{1}f_{l}=Q_{l}f_{l-1}+Q_{l+1}f_{l+1}\end{equation*}
\begin{equation*}\mathcal{C}_{2}f_{l}=Q_{l}Q_{l-1}f_{l-2}+(Q^2_{l+1}+Q^2_l)f_{l}+Q_{l+1}Q_{l+2}f_{l+2}\end{equation*}
\begin{equation*}\mathcal{S}_{1}f_{l}=(l-1)Q_lf_{l-1}-(l+2)Q_{l+1}f_{l+1}\end{equation*}
\begin{align*}\mathcal{S}_{2}f_{l}&=(l-2)Q_{l-1}Q_{l}f_{l-2}+(lQ_{l+1}^2-(l+1)Q_{l}^2)f_{l}\\&\quad-(l+3)Q_{l+2}Q_{l+1}f_{l+2}\end{align*}
\begin{equation*}\bar{S}_1f_l=lQ_{l+1}f_{l+1}-(l+1)Q_{l}f_{l-1}\end{equation*}
\begin{align*}\bar{S}_2f_l&=-(l+1)Q_lQ_{l-1}f_{l-2}+(lQ^2_{l+1}-(l+1)Q^2_l)f_l\\
&\quad+lQ_{l+1}Q_{l+2}f_{l+2}\end{align*}
\end{flushleft}
\subsection{Group 2}
\begin{equation*}A_{0}f_{l}=l(l+1)f_l\end{equation*}
\begin{equation*}A_{1}f_{l}=(l^2-1)Q_lf_{l-1}+l(l+2)Q_{l+1}f_{l+1}\end{equation*}
\begin{align*}A_{2}f_{l}&=(l+1)(l-2)Q_lQ_{l-1}f_{l-2}+l(l+3)Q_{l+1}Q_{l+2}f_{l+2}\\
&\quad+(Q_{l+1}^2l(l+3)+Q_l^2(l+1)(l-2))f_l\end{align*}
\begin{equation*}B_0f_{l}=0\end{equation*}
\begin{equation*}B_1f_l=imf_l\end{equation*}
\begin{equation*}B_2f_l=2im\big(Q_lf_{l-1}+Q_{l+1}f_{l+1}\big)\end{equation*}
\begin{equation*}\mathcal{X}_0f_{l}=im(l-1)(l+2)f_l\end{equation*}
\begin{equation*}\mathcal{X}_1f_{l}=im(l-2)(l+3)(Q_lf_{l-1}+Q_{l+1}f_{l+1})\end{equation*}
\begin{equation*}\bar{\mathcal{X}}_0f_{l}=(l-2)(l-1)(l+1)f_{l-1}Q_l-l(l+2)(l+3)Q_{l+1}f_{l+1}\end{equation*}
\begin{align*}
    \bar{\mathcal{X}}_1f_{l}&=Q_lQ_{l-1}f_{l-2}(l+1)(l-2)(l-3)\\
    &\quad-Q_{l+1}Q_{l+2}f_{l+2}l(l+3)(l+4)\\
    &\quad+f_l\Big[2m^2-Q_{l+1}^2[l^2(l+1)+2l(l+1)+4l]\\
    &\quad+Q_l^2[l(l+1)^2-2l(l+1)+4(l+1)]\Big]
\end{align*}
\begin{align*}\tilde{\mathcal{A}}_2f_l&=-(l-2)(l+1)Q_lQ_{l-1}f_{l-2}\\
&\quad+(l^2 Q_{l+1}^2+(l+1)^2Q_{l}^2-m^2)f_{l}\\
&\quad-l(l+3) Q_{l+1}Q_{l+2}f_{l+2}
\end{align*}
\begin{equation*}
\tilde{\mathcal{B}}_2f_l=2im\Big((l+1)Q_{l+1}f_{l+1}-lQ_lf_{l-1}\Big)
\end{equation*}
\begin{equation*}\bar{\mathcal{X}}f_l=2im\Big(Q_{l}f_{l-1}(l-2)-Q_{l+1}f_{l+1}(l+3)\Big)\end{equation*}
\begin{align*}\bar{\mathcal{W}}f_l&=Q_{l+1}Q_{l+2}f_{l+2}(l+3)(l+4)\\
&\quad+Q_{l}Q_{l-1}f_{l-2}(l-2)(l-3)\\
&\quad+f_{l}\Big(Q_{l+1}^2l(l-1)+Q_{l}^2(l+1)(l+2)\\
&\quad+2m^2-l(l+1)\Big)\end{align*}
\subsection{Group 3}

\begin{equation*}\mathcal{F}_0f_l=-im(l^2+l-2)f_l\end{equation*}
\begin{equation*}\mathcal{F}_1f_l=-im\Big((l-1)(l+4)Q_{l+1}f_{l+1}+(l-3)(l+2)Q_{l}f_{l-1}\Big)\end{equation*}
\begin{equation*}\mathcal{G}_0f_l=l(l-1)(l+2)Q_{l+1}f_{l+1}-(l-1)(l+1)(l+2)Q_{l}f_{l-1}\end{equation*}
\begin{align*}\mathcal{G}_1f_l&=Q_{l+1}Q_{l+2}f_{l+2}(l-1)l(l+3)\\
&\quad-Q_{l-1}Q_{l}f_{l-2}(l+1)(l-2)(l+2)\\
&\quad+f_l\Big(Q_{l}^2(l+1)(l^2-l+4)\\
&\quad-Q_{l+1}^2l(l^2+3l+6)
+2m^2\Big)\end{align*}
\begin{align*}\mathcal{W}f_l&=Q_{l+1}Q_{l+2}f_{l+2}l(l-1)\\
&\quad+Q_{l}Q_{l-1}f_{l-2}(l+1)(l+2)\\
&\quad+f_{l}\Big(Q_{l+1}^2l(l-1)+Q_{l}^2(l+1)(l+2)\\
&\quad+2m^2-l(l+1)\Big)\end{align*}
\begin{equation*}\mathcal{X}f_l=2im\Big(Q_{l}f_{l-1}(l+2)-Q_{l+1}f_{l+1}(l-1)\Big)f_{l}\end{equation*}
\begin{equation*}\mathcal{J}_0f_l=l(l^2-1)(l+2)f_{l}\end{equation*}
\begin{align*}\mathcal{J}_1f_l&=(l+2)(l+3)l(l-1)Q_{l+1}f_{l+1}\\
&\quad+(l-1)(l-2)(l+1)(l+2)Q_{l}f_{l-1}\end{align*}
\begin{align*}
    \mathcal{J}_2f_l&=(l+1)(l+2)(l-2)(l-3)Q_{l-1}Q_{l}f_{l-2}\\
    &\quad+l(l-1)(l+3)(l+4)Q_{l+1}Q_{l+2}f_{l+2}\\
    &\quad+\Big((l^4+10l^3+7l^2-18l)Q_{l+1}^2\\
    &\quad+(l^4-6l^3-17l^2+6l+16)Q_{l}^2\Big)f_{l}
\end{align*}
\begin{equation*}\mathcal{K}_0f_l=0\end{equation*}
\begin{equation*}\mathcal{K}_1f_l=-2im(l-1)(l+2)f_{l}\end{equation*}
\begin{equation*}\mathcal{K}_2f_l=-4im\Big[(l+3)(l-1)Q_{l+1}f_{l+1}+(l^2-4)Q_lf_{l-1}\Big]\end{equation*}
\begin{align*}\mathcal{T}f_l&=2im\big[(l-1)l(l+3)Q_{l+1}f_{l+1}\\
&\quad-(l-2)(l+1)(l+2)Q_{l}f_{l-1}\big]\end{align*}
\begin{align*}\bar{\mathcal{T}}f_l&=\Big((l^5+4l^4+7l^3-12l)Q_{l+1}^2\\
&\quad-(l^5+l^4+l^3+7l^2-2l-8)Q_{l}^2\Big)f_{l}\\
&\quad-(l-1)l(l+1)(l+3)(l+4)Q_{l+1}Q_{l+2}f_{l+2}\\
&\quad+(l-3)(l-2)l(l+1)(l+2)Q_{l-1}Q_{l}f_{l-2}
\end{align*}
\begin{align*}\mathcal{S}f_l&=\big(2l^5+3l^4-6l^3-5l^2+6l\big)Q_{l+1}^2f_l\\
&\quad-\big(2l^5+7l^4+2l^3-11l^2-4l+4\big)Q_{l}^2f_l\end{align*}
\begin{align*}\bar{\mathcal{S}}f_l&=im\big[(l-1)l(l+2)(l+3)Q_{l+1}f_{l+1}\\
&\quad+(l-2)(l-1)(l+1)(l+2)Q_lf_{l-1}\big]\end{align*}

%

\end{document}